\documentclass[reprint,aps,prd,onecolumn,notitlepage,showpacs,nofootinbib,preprintnumbers,superscriptaddress]{revtex4-1}
\usepackage{amsmath}
\usepackage{amsfonts}
\usepackage{amssymb}
\usepackage{latexsym}
\usepackage{color}
\usepackage{graphicx}
\usepackage{bm}
\usepackage{epsfig}
\usepackage[english]{babel}
\usepackage{times}
\usepackage{comment}
\usepackage{slashbox}

\def\in{\mathrm{in}}
\def\out{\mathrm{out}}
\def\ext{\mathrm{ext}}
\def\gs{\mathrm{gs}}
\def\CL{\mathrm{CL}}
\def\CV{\mathrm{CV}}
\def\CB{\mathrm{CB}}

\begin{document}

\title{Maximal volume behind horizons without curvature singularity}
\author{Shao-Jun Wang}
\author{Xin-Xuan Guo}
\author{Towe Wang}
\email[Electronic address: ]{twang@phy.ecnu.edu.cn}
\affiliation{Department of Physics, East China Normal University,\\
Shanghai 200241, China\\ \vspace{0.2cm}}
\date{\today\\ \vspace{1cm}}
\begin{abstract}
The black hole information paradox is related to the area of event horizon, and potentially to the volume and singularity behind it. One example is the complexity/volume duality conjectured by Stanford and Susskind. Accepting the proposal of Christodoulou and Rovelli, we calculate the maximal volume inside regular black holes, which are free of curvature singularity, in asymptotically flat and anti-de Sitter spacetimes respectively. The complexity/volume duality is then applied to anti-de Sitter regular black holes. We also present an analytical expression for the maximal volume outside the de Sitter horizon.
\end{abstract}


\maketitle




\section{Introduction}\label{sect-intro}
Information is obscured behind the horizon as reflected in the black hole information paradox \cite{Chen:2014jwq,Lochan:2016nbs}. There have been several solutions to this paradox, and there will be more until the fulfilment of the quantum theory of gravity. In most solutions to the paradox, one should take into consideration of the area, volume and singularity of black holes. For example, in reference \cite{Stanford:2014jda}, Stanford and Susskind proposed that the complexity of a state on the anti-de Sitter (AdS) boundary is proportional to the maximal spatial volume of the Einstein-Rosen bridge anchored at the boundary state. This refines Susskind's earlier conjecture \cite{Susskind:2014rva} that the complexity is proportional to the length of the Einstein-Rosen bridge. And it is later superseded by an elaborated conjecture \cite{Brown:2015bva,Brown:2015lvg} relating the complexity to the action of a Wheeler-DeWitt patch. As argued with strong evidences in references \cite{Susskind:2014rva,Stanford:2014jda}, even after the boundary field theory reaches ordinary local equilibrium at the scrambling time, subtle quantum properties continue to equilibrate until the recurrence time, and the computational complexity of the boundary state proceeds to increase from $S\ln S$ to $e^{S}$ between the two times. Here $S$ stands for the Bekenstein-Hawking entropy.

Apparently independent of the above stream, Christodoulou and Rovelli define the volume of a black hole as the volume of maximal spacelike hypersurface bounded by the event horizon \cite{Christodoulou:2014yia}. This was initially illustrated in detail with spherically symmetric spacetime, such as the Schwarzschild and the Reissner-Nordstr\"{o}m (RN) black holes. Later, more efforts were made on the extension to the Kerr black hole \cite{Bengtsson:2015zda}, the relations between volume and entropy \cite{Ong:2015tua,Zhang:2015gda,Astuti:2016dmk}, and the effects of Hawking radiation \cite{Ong:2015dja,Christodoulou:2016tuu,Bhaumik:2016sav}.

So far the main attention has been paid on interior volume of black holes with curvature singularities, except for a very recent work \cite{Zhang:2016sjy} which turned to non-commutative black holes. As emphasized by references \cite{Ong:2016xcq,Ong:2016iwi}, to resolve the information paradox, despite the potentially important role of volumes, we should not ignore the singularity. In the present paper, we will explore the case without curvature singularity, specifically two examples: the volume inside regular black holes in sections \ref{sect-rbh} and \ref{sect-AdS}, and the volume outside the de Sitter (dS) horizon in section \ref{sect-dS}.

Although developed independently, the black hole volume defined by Christodoulou and Rovelli ought to be useful for the complexity/volume duality in the infinite-time limit \cite{Stanford:2014jda}, as we will emphasize in section \ref{sect-AdS}. Remember that the complexity is defined for a state on the AdS boundary. Therefore, to apply the complexity/volume duality, we should restrict our investigations to black holes in asymptotically AdS spacetime. But in section \ref{sect-rbh} the examples collected from literature \cite{bardeen68,AyonBeato:2000zs,Hayward:2005gi,AyonBeato:1999rg,Dymnikova:2004zc,AyonBeato:1998ub} are asymptotically flat at spatial infinity. Fortunately, an asymptotically AdS regular black hole was constructed recently by Fan in reference \cite{Fan:2016rih}, see more solutions in reference \cite{Fan:2016hvf}. Partially inspired by his work, in appendix \ref{app-cc}, we will demonstrate a prescription to switch on cosmological constant in a class of solutions for the Einstein gravity. This prescription can be applied to all solutions in section \ref{sect-rbh} to get their AdS counterparts. In section \ref{sect-AdS}, to such asymptotically AdS solutions we will apply Christodoulou and Rovelli's definition of black hole volume together with Stanford and Susskind's conjecture of complexity/volume duality. Some loose ends will be discussed in section \ref{sect-disc}. In appendix \ref{app-ext}, we will work out the extremal AdS regular black holes as the ground state of complexity growth rate.

\section{Volume inside regular black holes}\label{sect-rbh}
Throughout this paper, we will work in units in which $G_N=\hbar=c=1$, and restrict our study to $(3+1)$-dimensional spherical static spacetimes described by the line element
\begin{equation}\label{let}
ds^2=-f(r)dt^2+\frac{1}{f(r)}dr^2+r^2d\Omega^2
\end{equation}
where $d\Omega^2=d\theta^2+\sin^2\theta d\phi^2$. Using the advanced time $v=t+\int f^{-1}(r)dr$, it can be rewritten as
\begin{equation}\label{lev}
ds^2=-f(r)dv^2+2dvdr+r^2d\Omega^2.
\end{equation}
A specific form of the lapse function $f(r)$ dictates a special spacetime. For instance, the RN spacetime is determined by a lapse function of the form
\begin{equation}\label{RN}
f(r)=1-\frac{2m}{r}+\frac{q^2}{r^2}.
\end{equation}
In the chargeless case $q=0$, it reduces to the Schwarzschild spacetime.

It is well known that both the Schwarzschild and the RN black holes have a curvature singularity at the spatial origin $r=0$ inside the event horizon. In contrast, regular black holes retain event horizons but get rid of curvature singularities. In reference \cite{AyonBeato:1998ub}, Ayon-Beato and Garcia showed in a specific example that regular black holes are solutions to Einstein equations coupled to nonlinear electrodynamics. Since then many solutions of regular black hole have been constructed in the literature.

\begin{table}[ht]
\begin{tabular}{l|c|c|c}
Lapse function & Extremality condition & Reference & Tag \\ \hline
$f(r)=1-\frac{2mr^2}{(r^2+q^2)^{3/2}}$ & $\frac{q}{m}\approx0.77$ & \cite{bardeen68,AyonBeato:2000zs} & RBHa\\
$f(r)=1-\frac{2mr^2}{r^3+q^3}$ & $\frac{q}{m}\approx1.06$ & \cite{Hayward:2005gi} & RBHb\\
$f(r)=1-\frac{2m}{r}\left(1-\tanh\frac{q^2}{2mr}\right)$ & $\frac{q}{m}\approx1.05$ & \cite{AyonBeato:1999rg} & RBHc\\
$f(r)=1-\frac{4m}{\pi r}\left(\arctan\frac{8mr}{\pi q^2}-\frac{8m\pi q^2r}{64m^2r^2+\pi^2q^4}\right)$ & $\frac{q}{m}\approx1.07$ & \cite{Dymnikova:2004zc} & RBHd\\
$f(r)=1-\frac{2mr^2}{(r^2+q^2)^{3/2}}+\frac{q^2r^2}{(r^2+q^2)^2}$ & $\frac{q}{m}\approx0.63$ & \cite{AyonBeato:1998ub} & RBHe\\
\end{tabular}
\caption{The lapse function of some regular black holes asymptotic to flat spacetime at $r\rightarrow\infty$. To facilitate comparison, we have transformed the model parameters into the mass $m$ and the charge $q$. Please refer to original references for details.}\label{tab-rbh}
\end{table}

For several regular black holes \cite{bardeen68,AyonBeato:2000zs,Hayward:2005gi,AyonBeato:1999rg,Dymnikova:2004zc,AyonBeato:1998ub}, the lapse functions are summarized in table \ref{tab-rbh}. At the spatial infinity $r\rightarrow\infty$, all of them behave as $f(r)\rightarrow1$, corresponding to asymptotically flat spacetime. There are two parameters: the mass $m$ and the charge $q$. As long as $q\neq0$, these black holes are free of singularity. When the charge-to-mass ratio is smaller than the extremal value in table \ref{tab-rbh}, each of them has two horizons $r=r_{\pm}$, which shrink into a single one under the extremality condition. In this section, we will compute the maximal volume of these black holes by following Christodoulou and Rovelli's recipe. As they demonstrated in reference \cite{Christodoulou:2014yia}, the maximal volume bounded by horizons of a spherical black hole can be evaluated via the integral
\begin{equation}\label{V}
V_{\max}=4\pi\int_{r_{-}}^{r_{+}}dr\frac{r^4}{\sqrt{A_{\min}^2+r^4f(r)}}+4\pi\int_0^{r_{-}}dr\frac{r^2}{\sqrt{f(r)}}.
\end{equation}
Here $A_{\min}$ is the maximum of $\sqrt{-r^4f(r)}$ in the range $r_{-}<r<r_{+}$. In other words, $A_{\min}^2$ is the minimal constant that guarantees the expression nonnegative under the radical sign.

In the expression \eqref{V}, the first term is often divergent. The second term, which will be denoted by $V_{\in}$, yields often a negligible contribution. One can interpret the second term as the volume bounded by the inner horizon, and the first term naively as the volume between inner and outer horizons. In most cases, the interval $(r_{-},r_{+})$ is finite, then the volume at large $v$, as was proven in reference \cite{Christodoulou:2014yia}, grows linearly with advanced time,
\begin{equation}\label{Vv}
V_{\max}\sim4\pi A_{\min}v.
\end{equation}
It is interesting to note that this relation implies the consistence between two conjectures in references \cite{Susskind:2014rva,Stanford:2014jda}, which state that the complexity is proportional to the length and the maximal volume of the Einstein-Rosen bridge respectively. Put into formulae, these conjectures claim that, after the scrambling and before the recurrence, the complexity of the boundary state can be estimated by $\mathcal{C}\sim TSd$ and $\mathcal{C}\sim V_{\max}/\ell_{c}$ respectively. Here $T$ represents the Hawking temperature, and $\ell_{c}$ is a length scale undetermined. Note that the length of the Einstein-Rosen bridge $d\sim t\sim v$ long after the scrambling, it is easy to see the two conjectures are consistent if $\ell_{c}\sim 4\pi A_{\min}/(TS)$. For small Schwarzschild-AdS black holes,\footnote{Recall that the Hawking temperature $T=f'(r_{+})/(4\pi)$, while the Bekenstein-Hawking entropy $S=\pi r_{+}^2$, with $r_{+}$ being the horizon radius and $f'(r_{+})=df(r_{+})/dr_{+}$. Limited to small Schwarzschild-AdS black holes, one has $r_{+}\sim2m$, $r_{+}f'(r_{+})\sim1$ and $V_{\max}\sim3\sqrt{3}\pi m^2v$ at large $v$.} we can get an estimate $\ell_{c}\sim3\sqrt{3}\pi r_{+}$, the same order as expected in reference \cite{Brown:2015lvg}.

\begin{figure}
\centering
\includegraphics[width=0.3\textwidth]{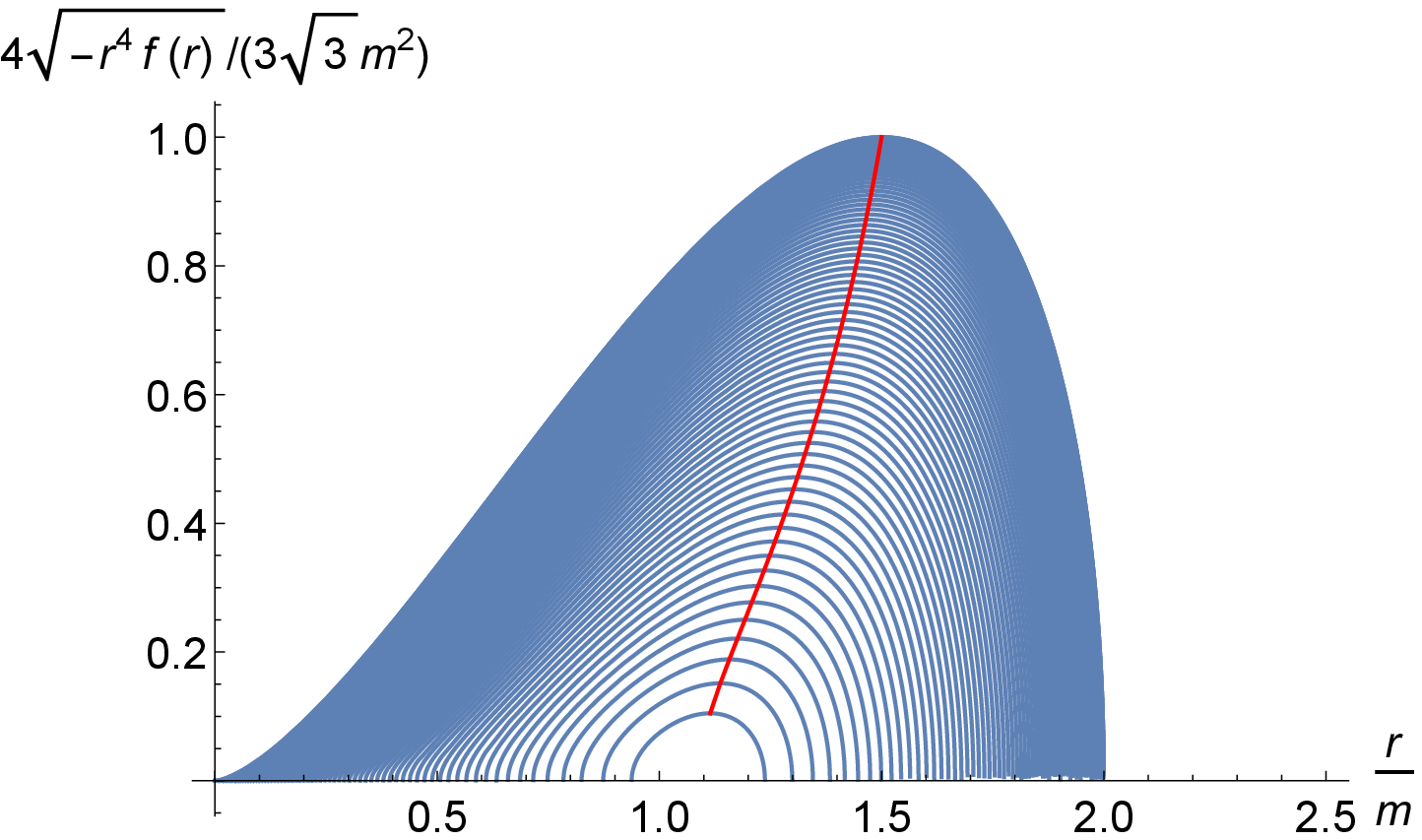}\includegraphics[width=0.3\textwidth]{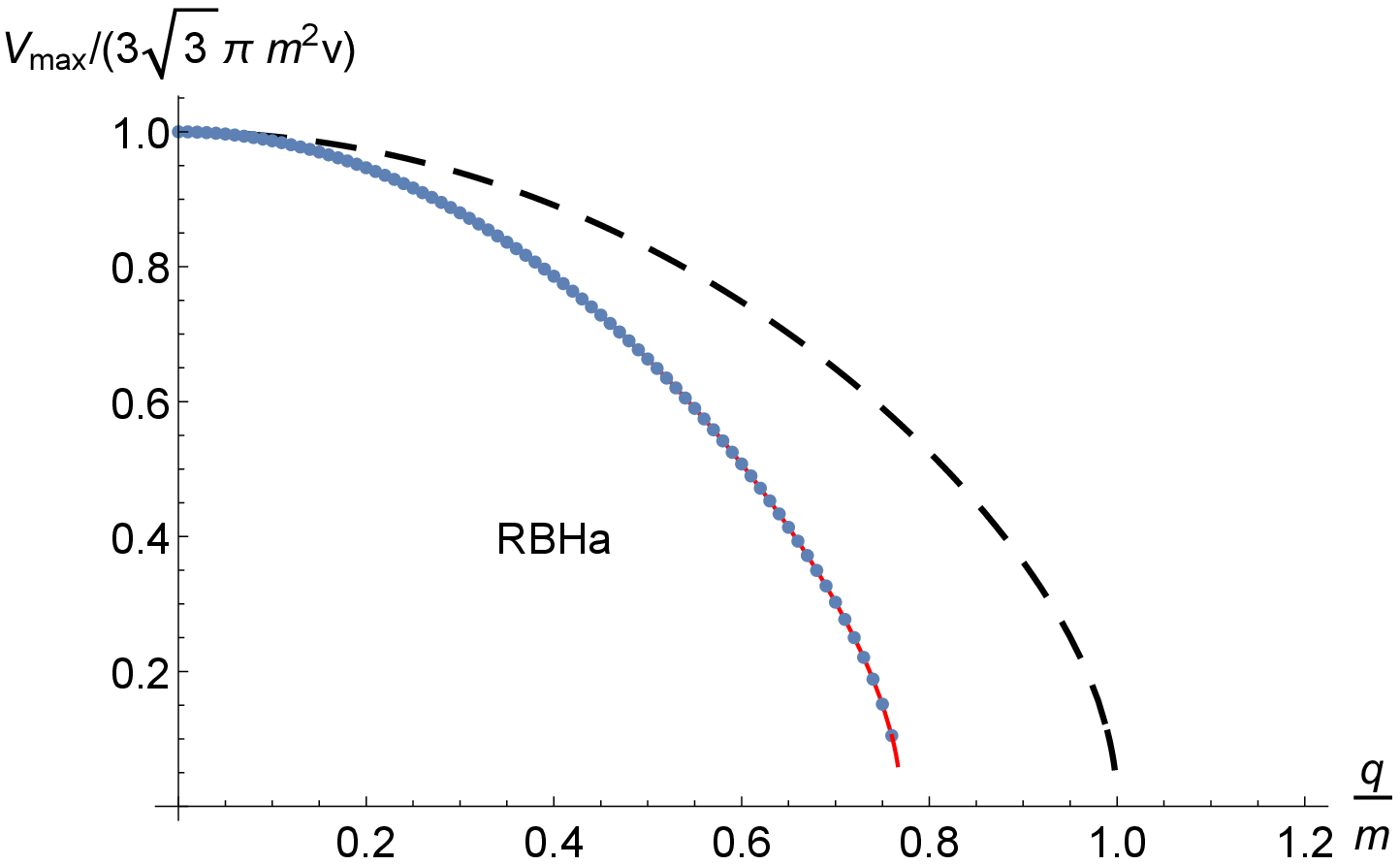}\includegraphics[width=0.3\textwidth]{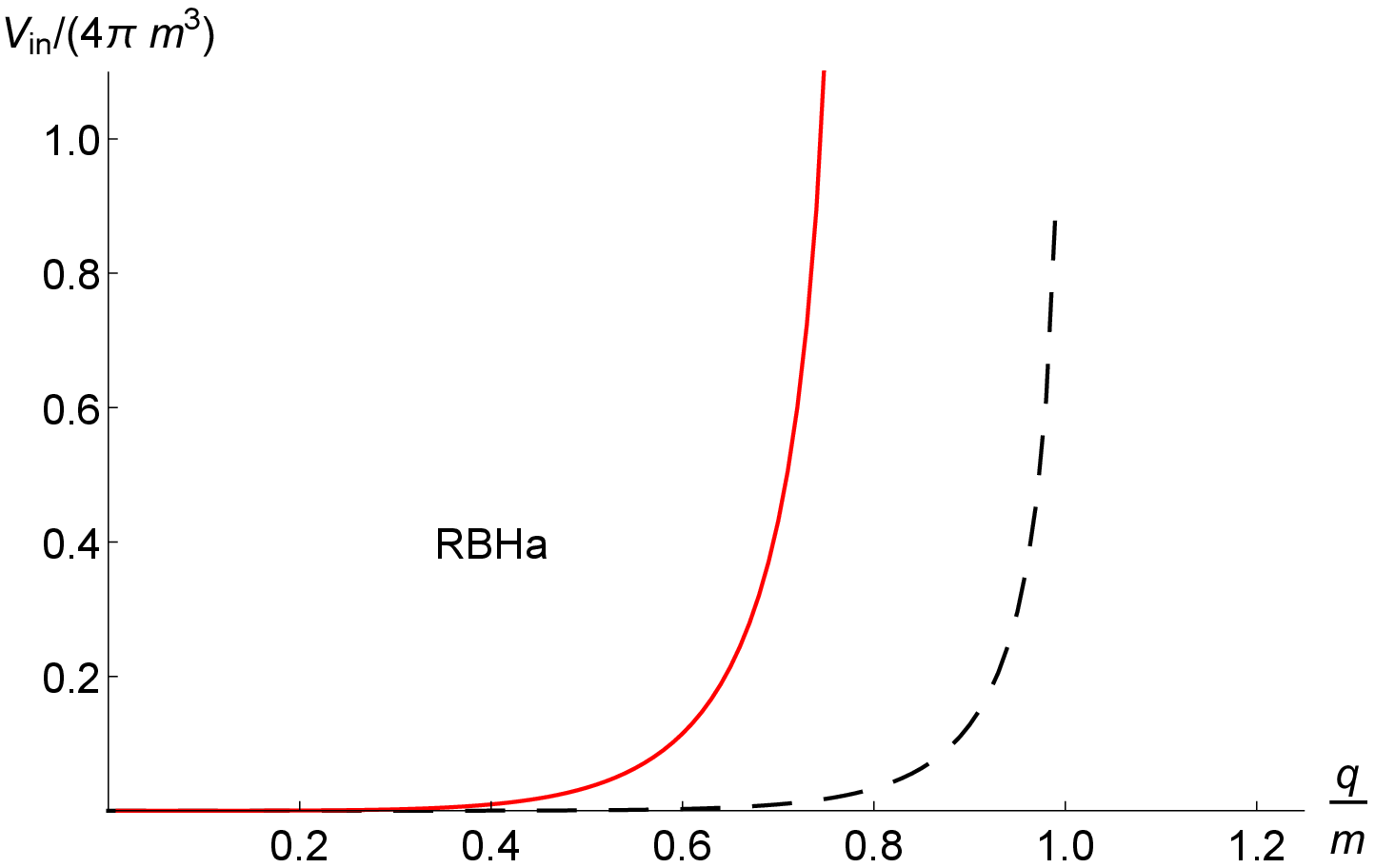}\\
\includegraphics[width=0.3\textwidth]{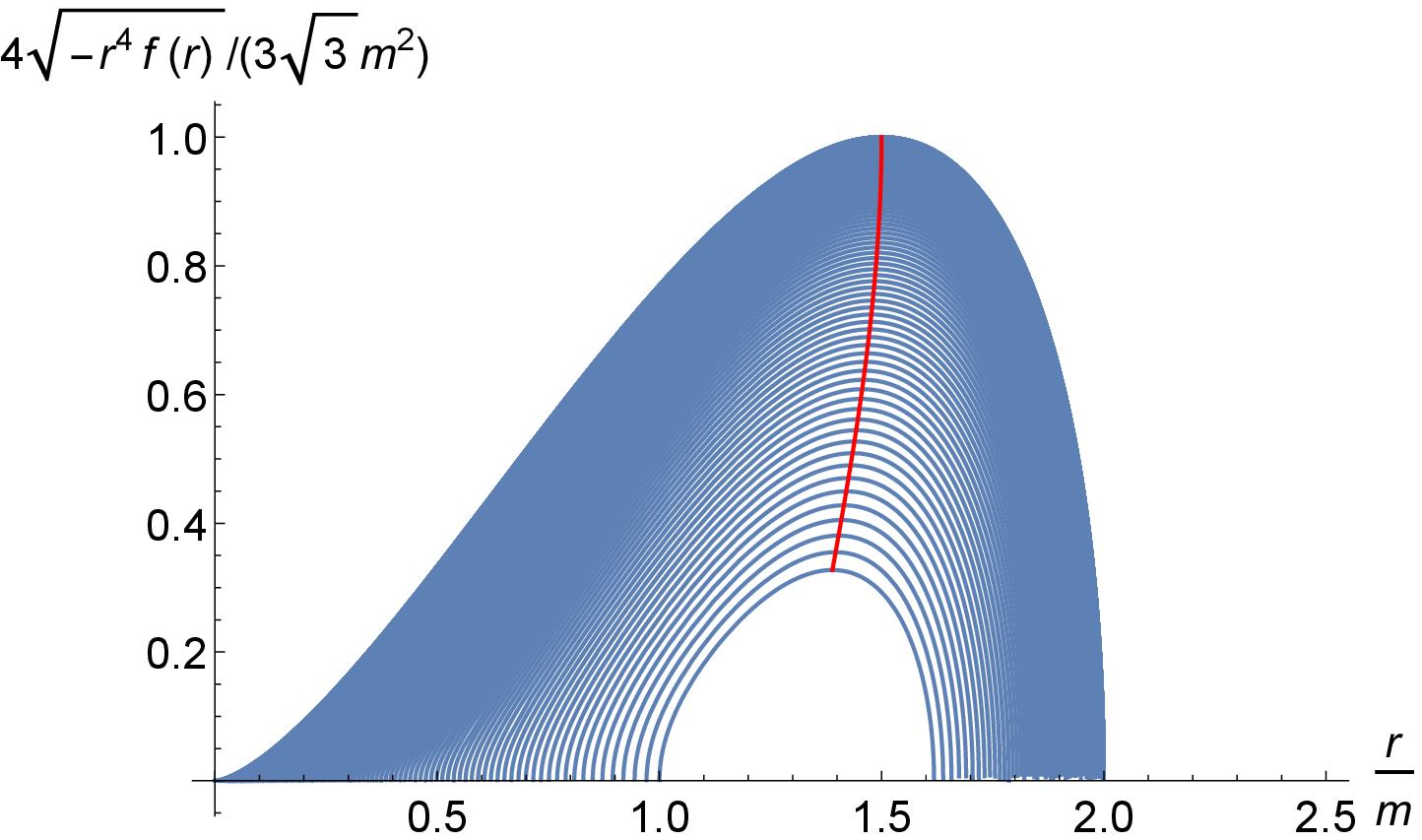}\includegraphics[width=0.3\textwidth]{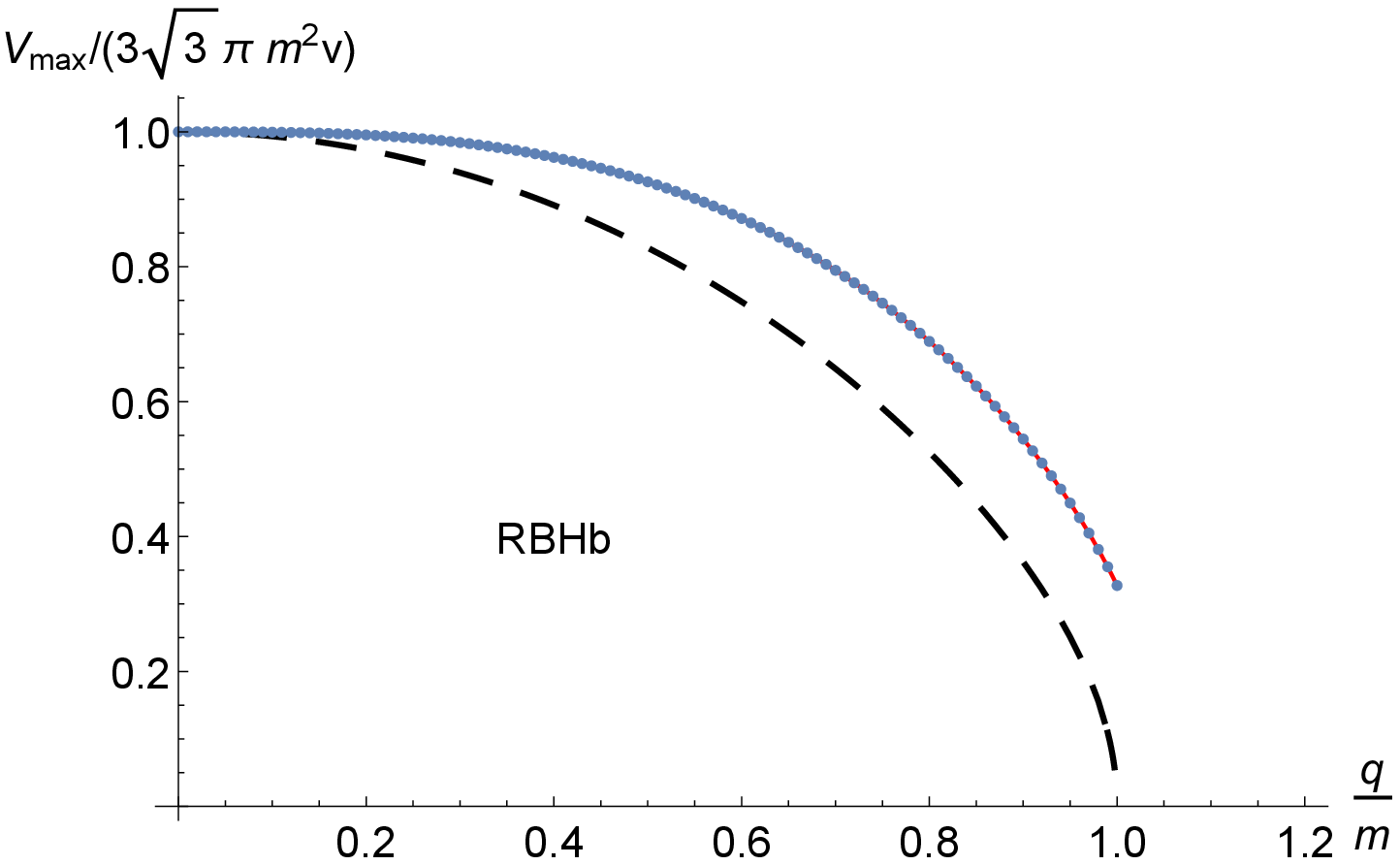}\includegraphics[width=0.3\textwidth]{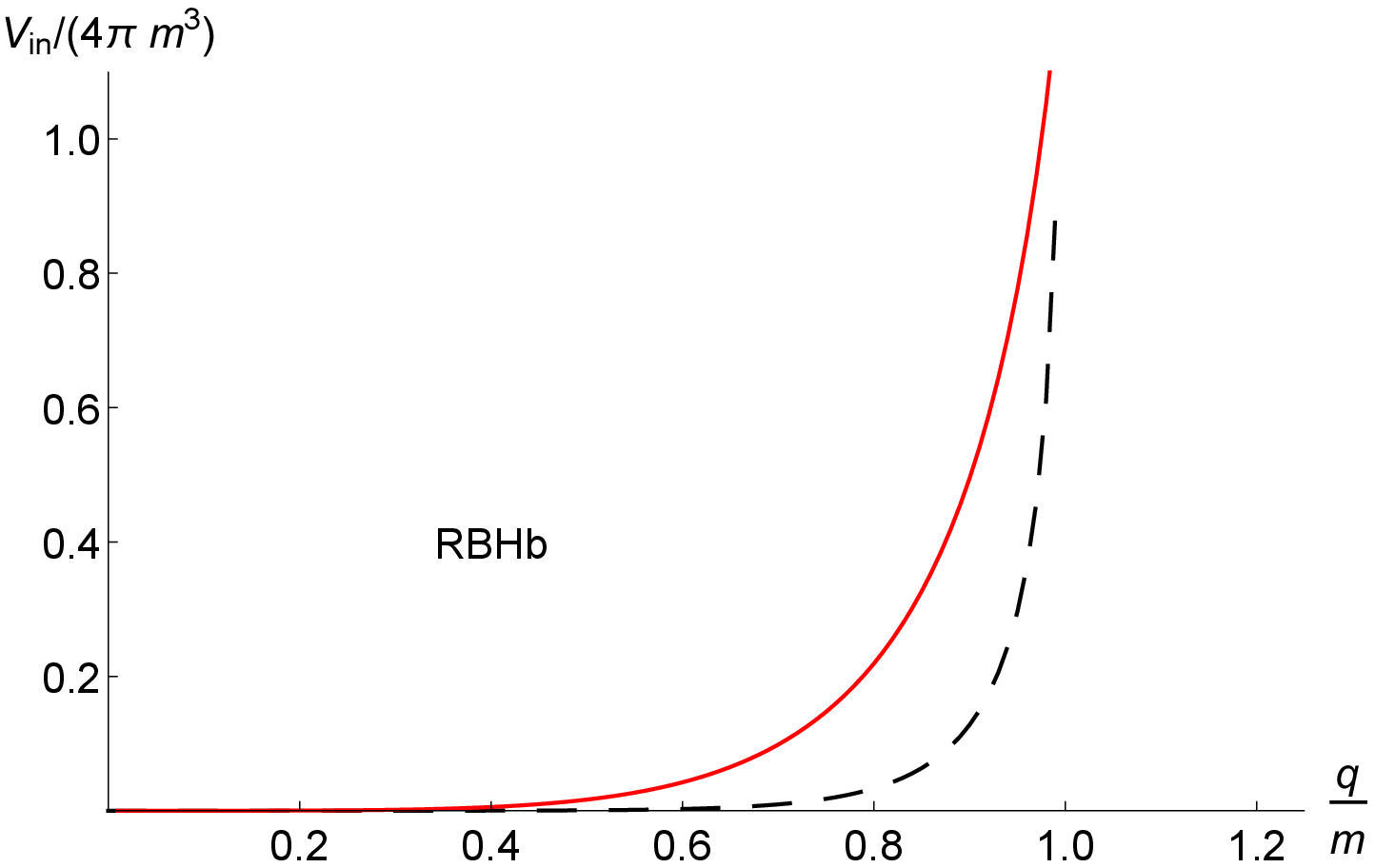}\\
\includegraphics[width=0.3\textwidth]{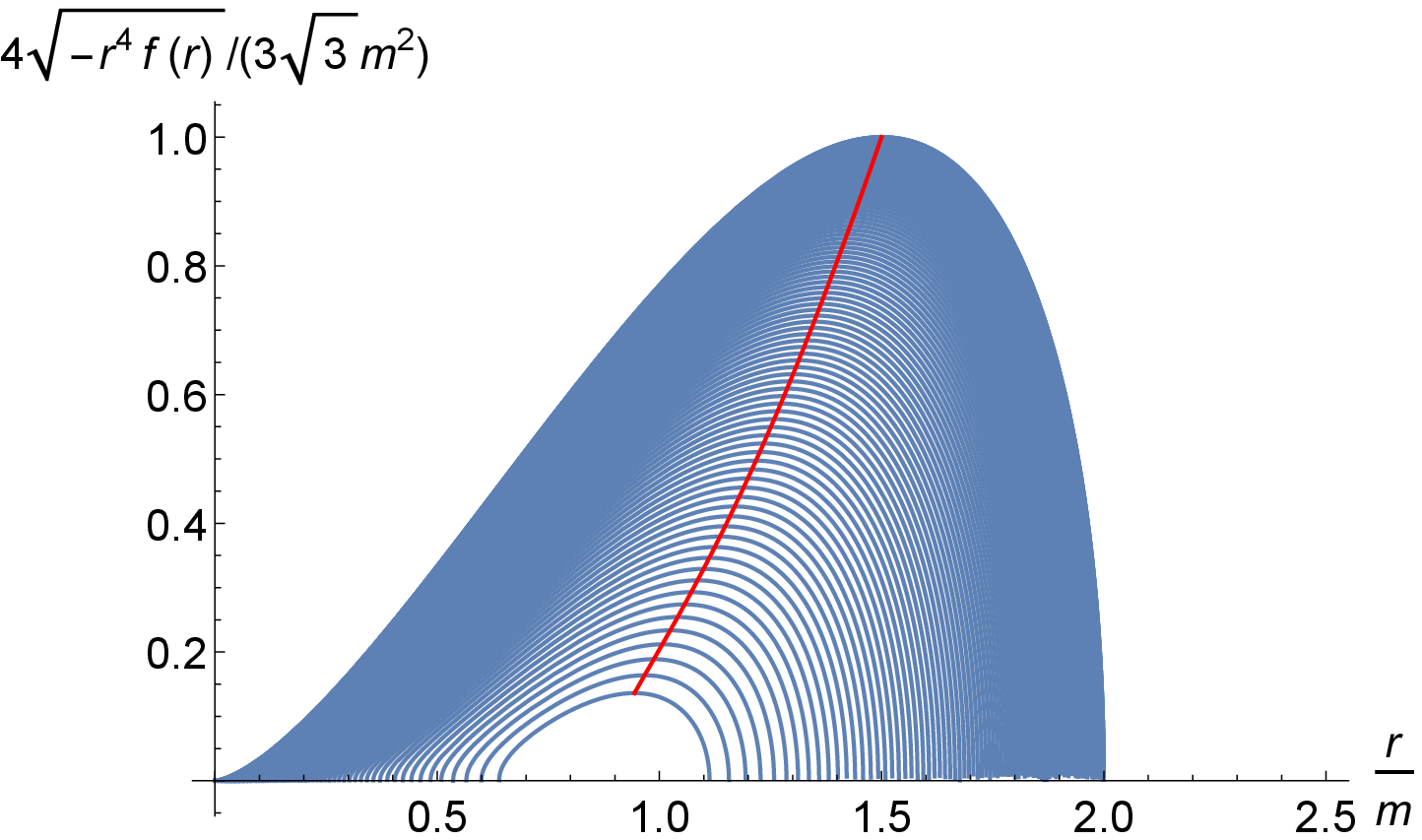}\includegraphics[width=0.3\textwidth]{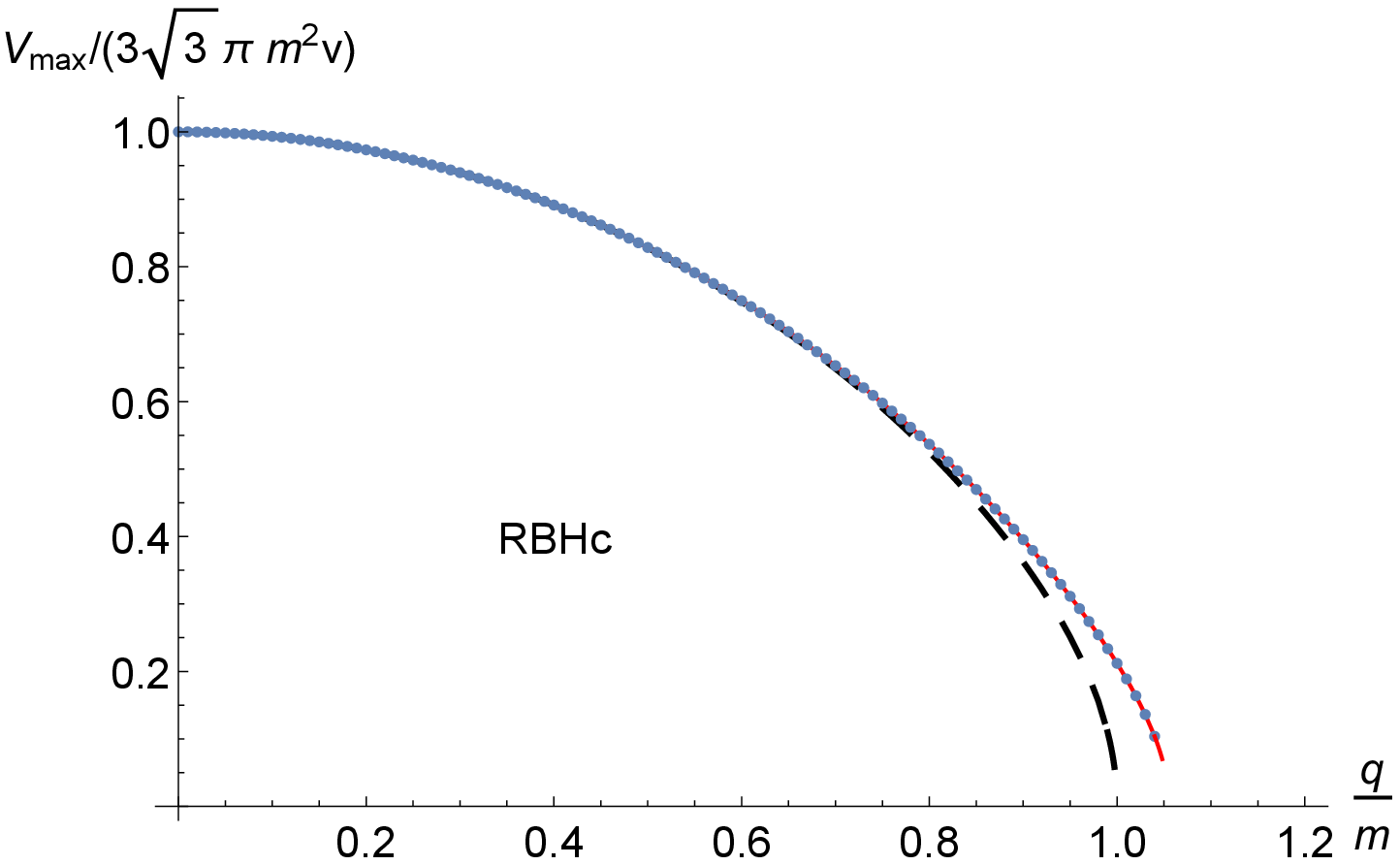}\includegraphics[width=0.3\textwidth]{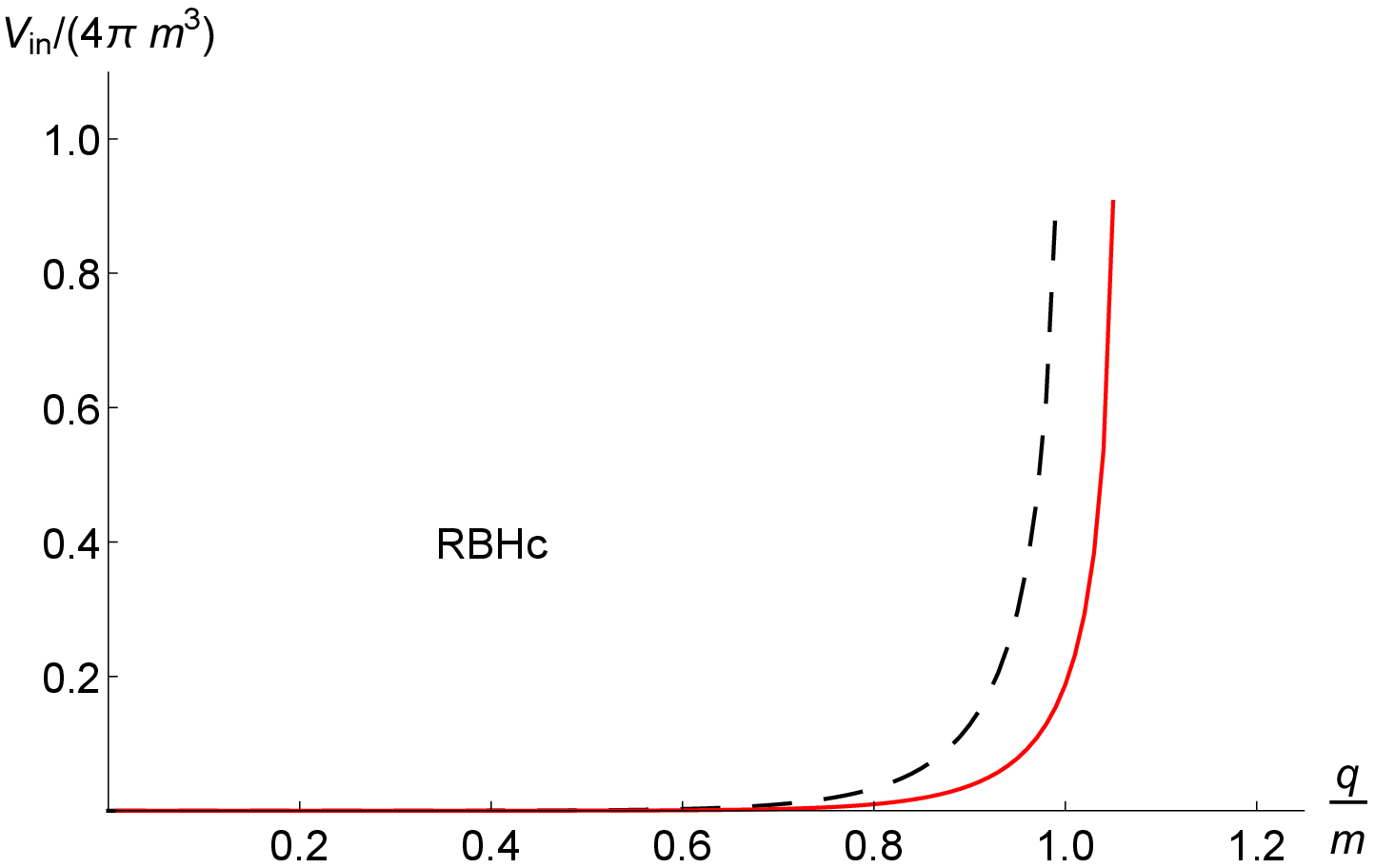}\\
\includegraphics[width=0.3\textwidth]{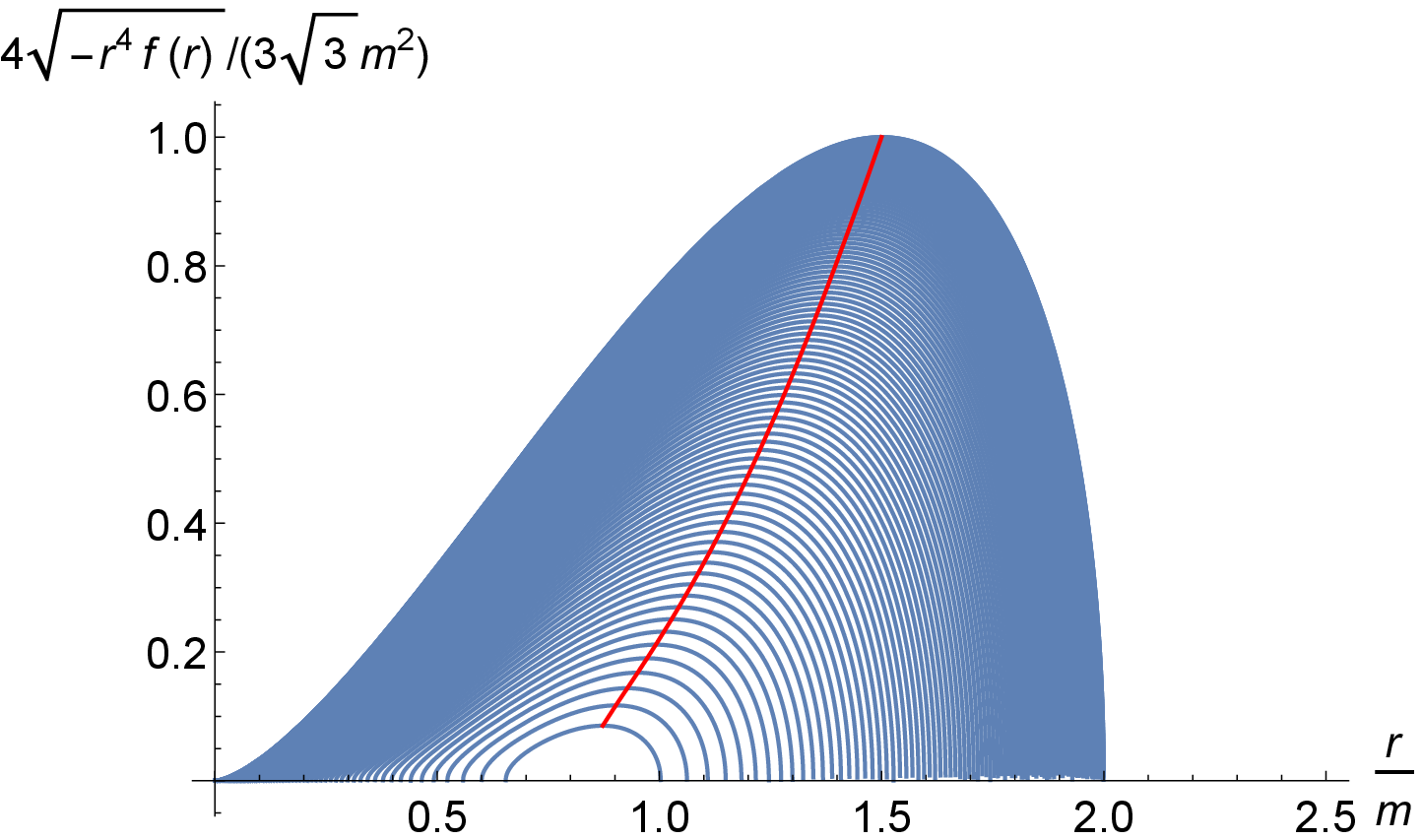}\includegraphics[width=0.3\textwidth]{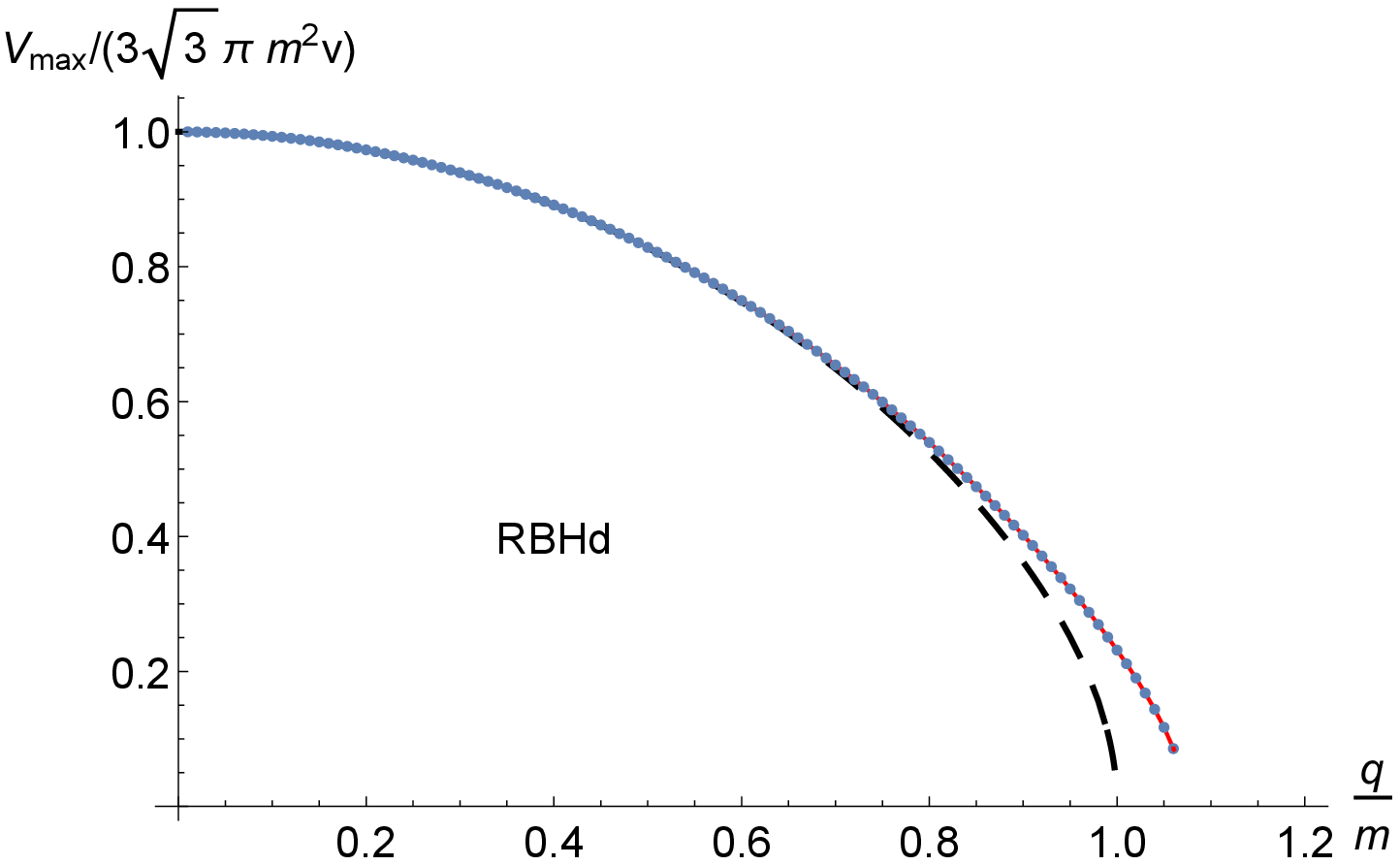}\includegraphics[width=0.3\textwidth]{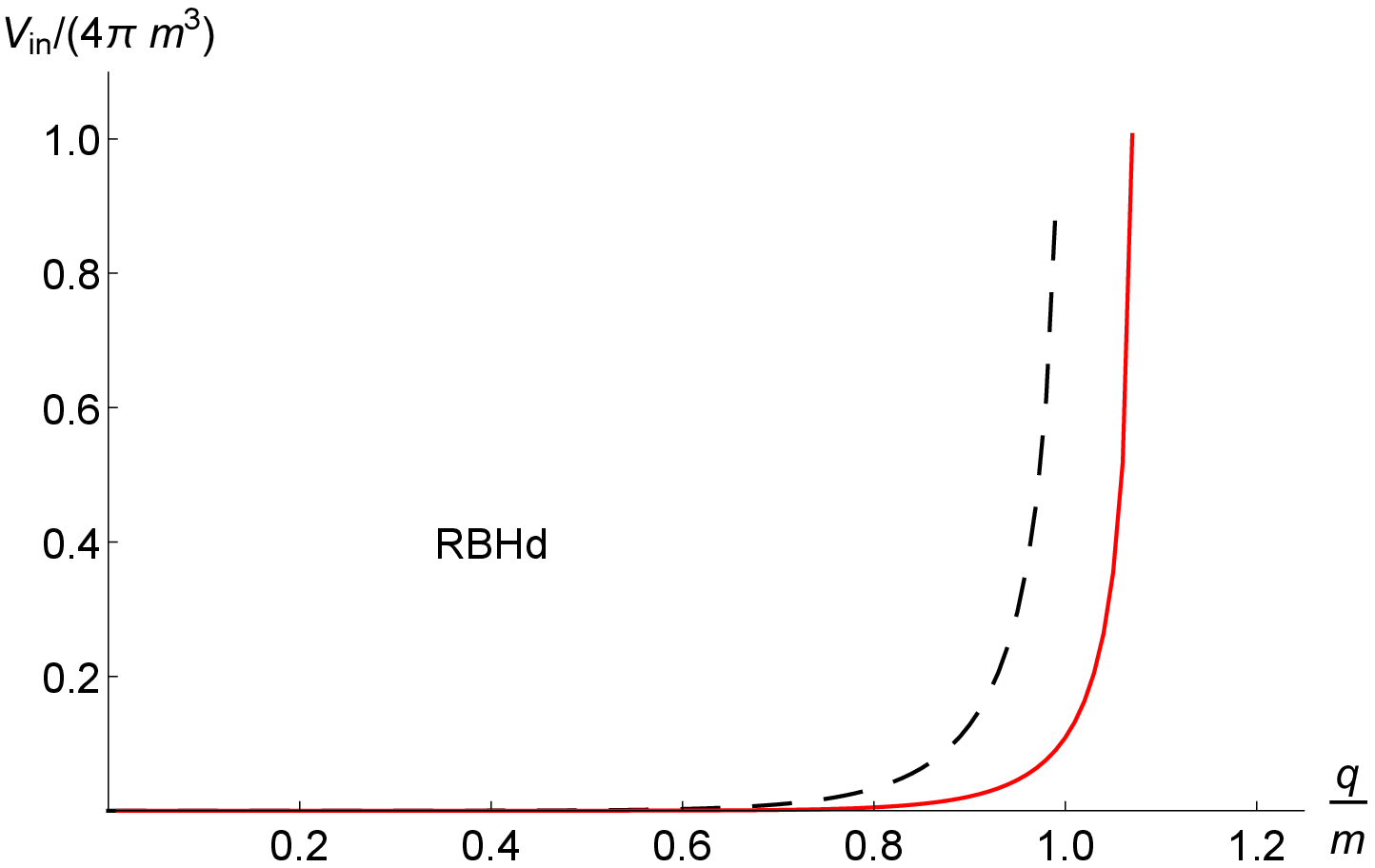}\\
\includegraphics[width=0.3\textwidth]{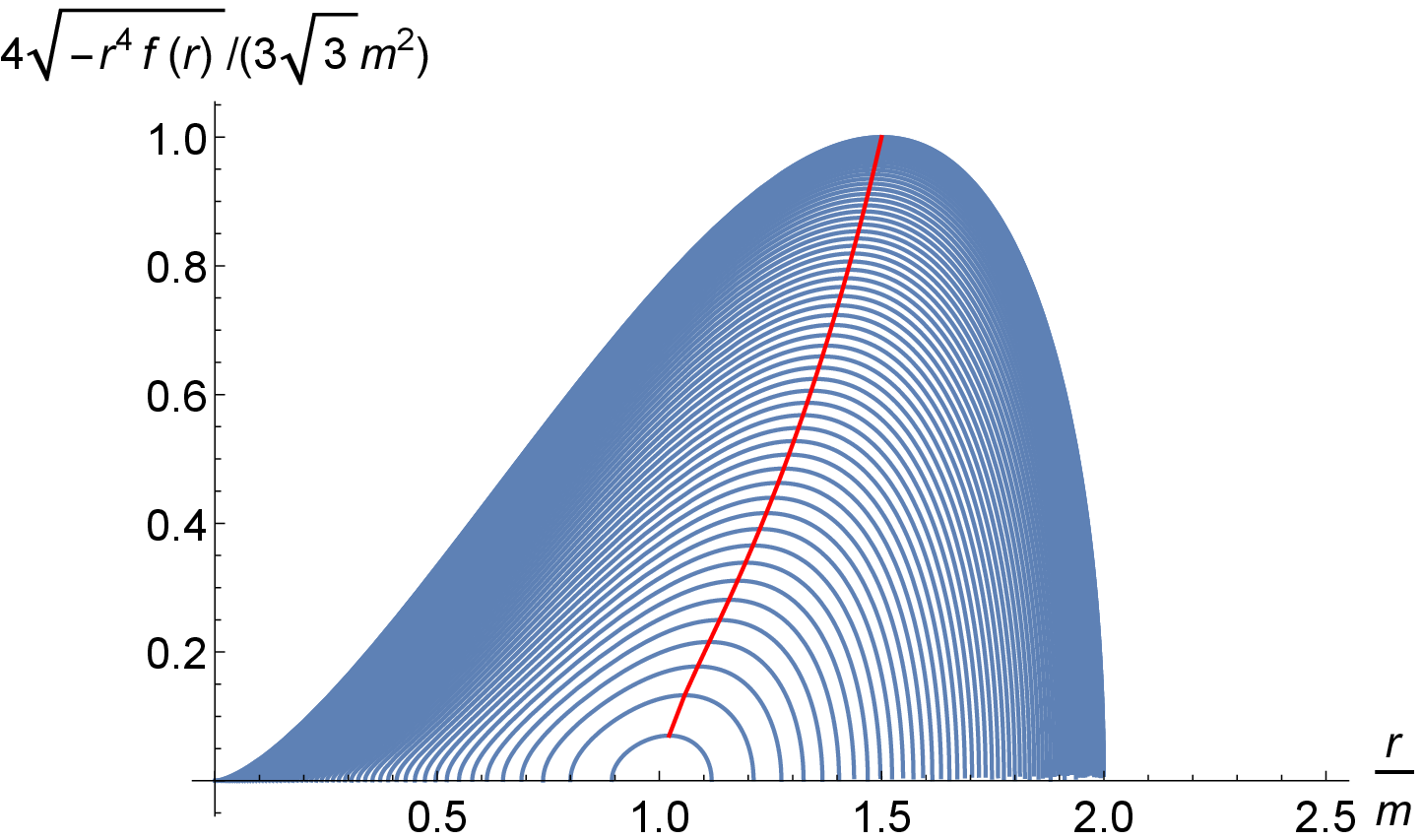}\includegraphics[width=0.3\textwidth]{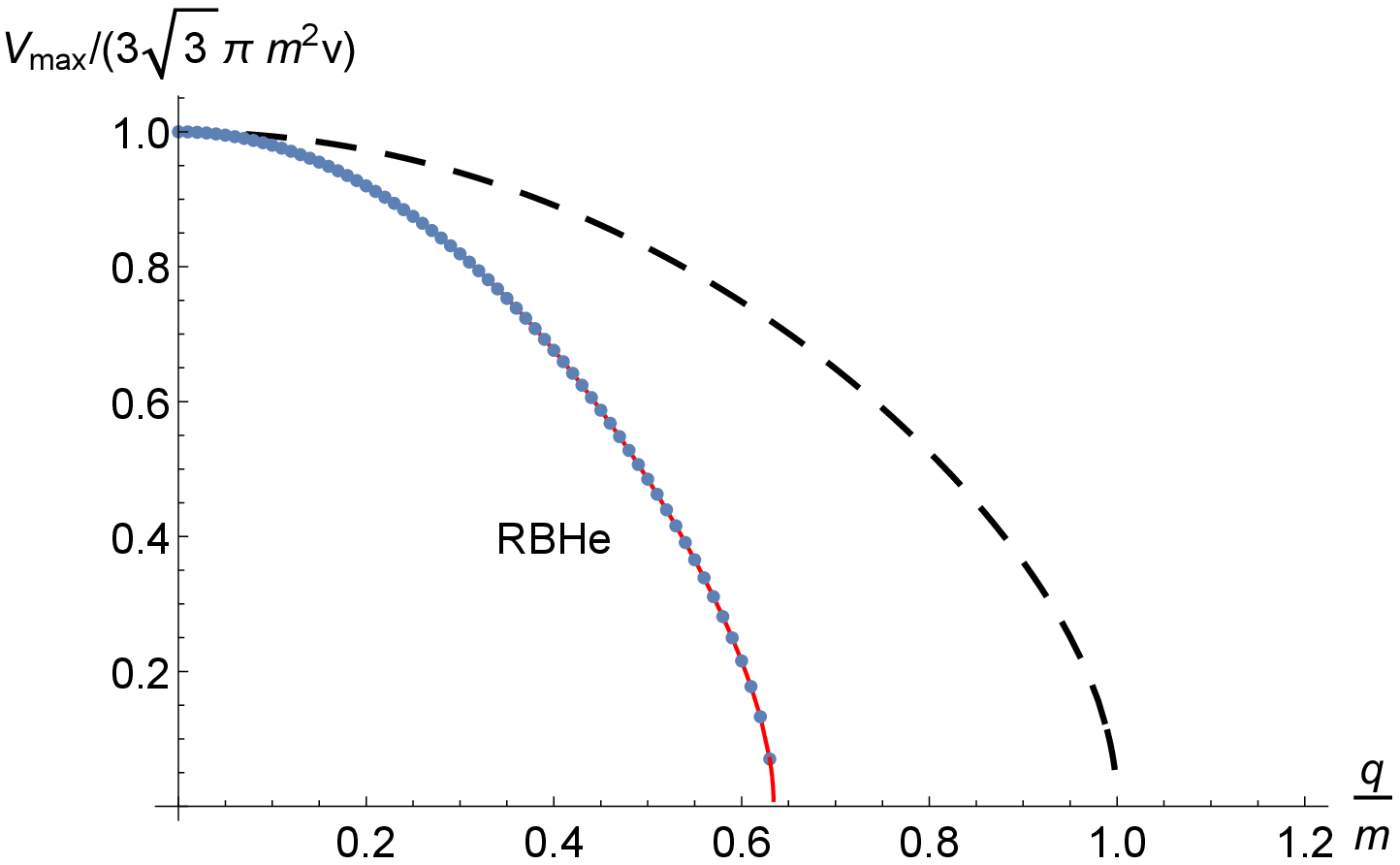}\includegraphics[width=0.3\textwidth]{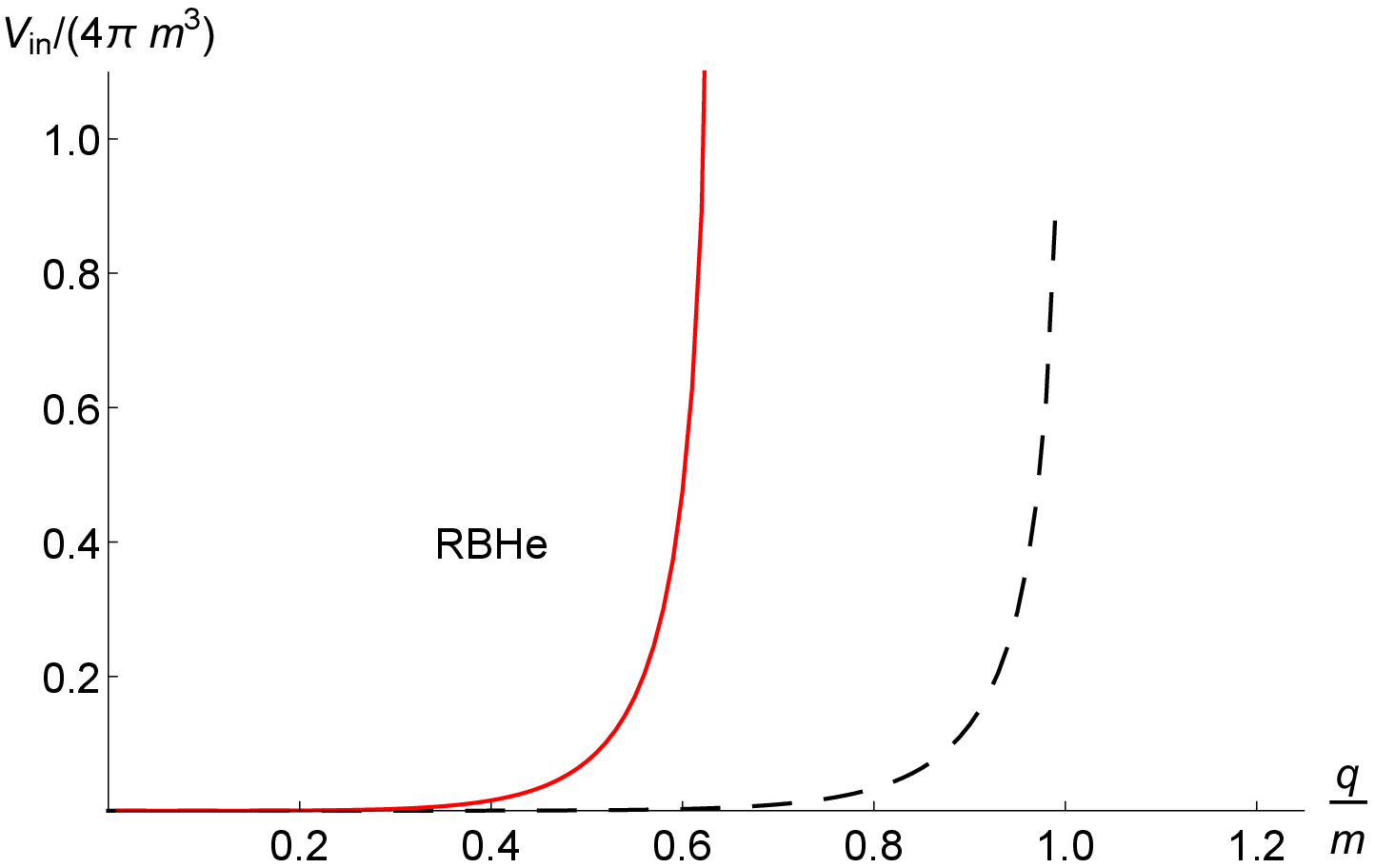}\\
\caption{(color online). Left and middle panels: the growth rate of maximal volume at large $v$, normalized by the Schwarzschild limit $3\sqrt{3}\pi m^2$. Right panels: the volume inside the inner horizon, normalized by $4\pi m^3$. Red and blue curves from top to bottom, tagged as RBHa, RBHb, RBHc, RBHd, RBHe accordingly, correspond to the black hole solutions in table \ref{tab-rbh} row by row. Black dashed lines depict results of the RN black hole for comparison.}\label{fig-rbh}
\end{figure}

For the Schwarzschild black hole \cite{Christodoulou:2014yia}, one finds $V_{\max}\sim3\sqrt{3}\pi m^2v$ at large $v$. For more complicated solutions, usually one cannot calculate $A_{\min}$ analytically, but it is feasible to numerically evaluate the function $\sqrt{-r^4f(r)}$ and seek for its maximum
\begin{equation}\label{Vf}
\frac{V_{\max}}{4\pi v}\sim\max\left[\sqrt{-r^4f(r)}\right].
\end{equation}
Using the parameter $m$ to non-dimensionalize all quantities and varying the other parameter $q$ from zero to the extremal value, we have performed the numerical evaluation for all of the black holes in table \ref{tab-rbh}. The results are shown in figure \ref{fig-rbh}.

To facilitate comparison, we tagged the five regular black hole solutions with RBHa, RBHb, etc in table \ref{tab-rbh} and figure \ref{fig-rbh} correspondingly. In every row of figure \ref{fig-rbh}, each value of $q$ gives a blue curve in the left panel and a blue point in the middle panel. As the charge $q$ varies from zero to its extremal value continuously, the maximum \eqref{Vf} decreases along the red trajectory in both left and middle panels. Therefore, according to Christodoulou and Rovelli's definition of black hole volume, we can say that, given the mass $m$ of a black hole, the less charge it has, the faster it grows in volume at large $v$. This behavior is similar to that of RN black holes, as depicted by black dashed lines in middle panels of figure \ref{fig-rbh}.

We end this section by some comments on the volume $V_{\in}$ bounded by the inner horizon. In the extremal case $r_{+}=r_{-}$, the first term of formula \eqref{V} vanishes and the second term becomes divergent. Then the leading-order contribution is
\begin{eqnarray}
\nonumber V_{\in}&=&4\pi\int_0^{r_{-}}dr\frac{r^2}{\sqrt{f(r)}}\\
\nonumber&\sim&\lim_{r_{+}\rightarrow r_{-}}4\pi\int_0^{r_{-}}dr\frac{r^2}{\sqrt{\frac{1}{2}f''(r_{-})(r_{+}-r)(r_{-}-r)}}\\
&\sim&\lim_{r_{+}\rightarrow r_{-}}-\frac{4\sqrt{2}\pi r_{-}^2}{\sqrt{f''(r_{-})}}\ln(r_{+}-r_{-}).
\end{eqnarray}
Hereafter primes denote derivatives with respect to the indicated variables. For non-extremal values of $q$, we have evaluated $V_{\in}$ numerically in right panels of figure \ref{fig-rbh}. In all examples, the curve of $V_{\in}$ becomes very steep as $q$ increases in the near-extremal region.


\section{AdS regular black holes and complexity/volume duality}\label{sect-AdS}
This section is motivated by the observation that relation \eqref{Vf} coincides with equation (2.5) in reference \cite{Stanford:2014jda},
\begin{equation}
\frac{dV}{d\tau}=\omega_{D-2}r^{D-2}\sqrt{|f(r)|},
\end{equation}
where $V$ is the volume of the infinite-time Einstein-Rosen bridge, and $\omega_{D-2}$ is the volume of a unit ($D-2$)-sphere. The coincidence occurs when we restrict Christodoulou and Rovelli's proposal to asymptotically AdS black holes and take the infinite-time limit of Stanford and Susskind's conjecture. We intend to put the two ingredients together in regular black holes, but there are two obstacles.

\begin{figure}
\centering
\includegraphics[width=0.3\textwidth]{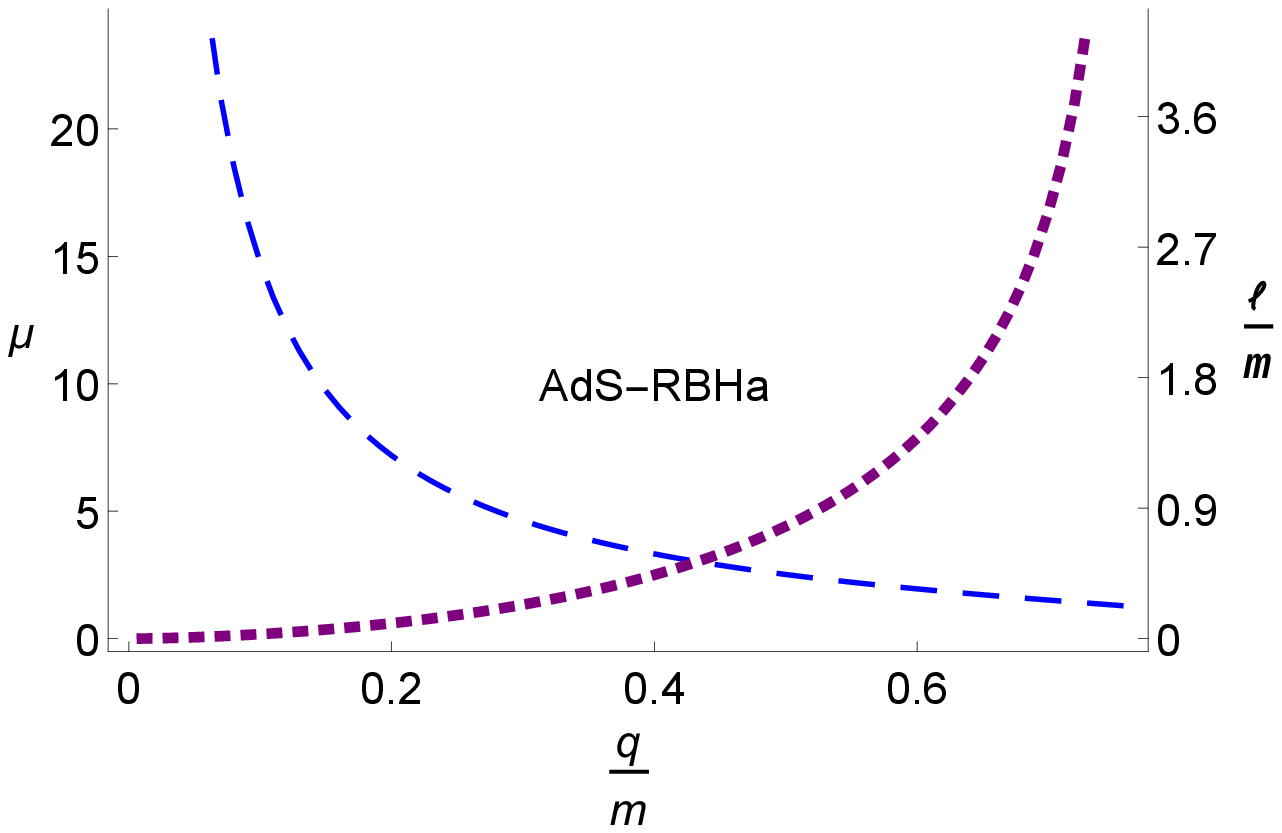}\includegraphics[width=0.3\textwidth]{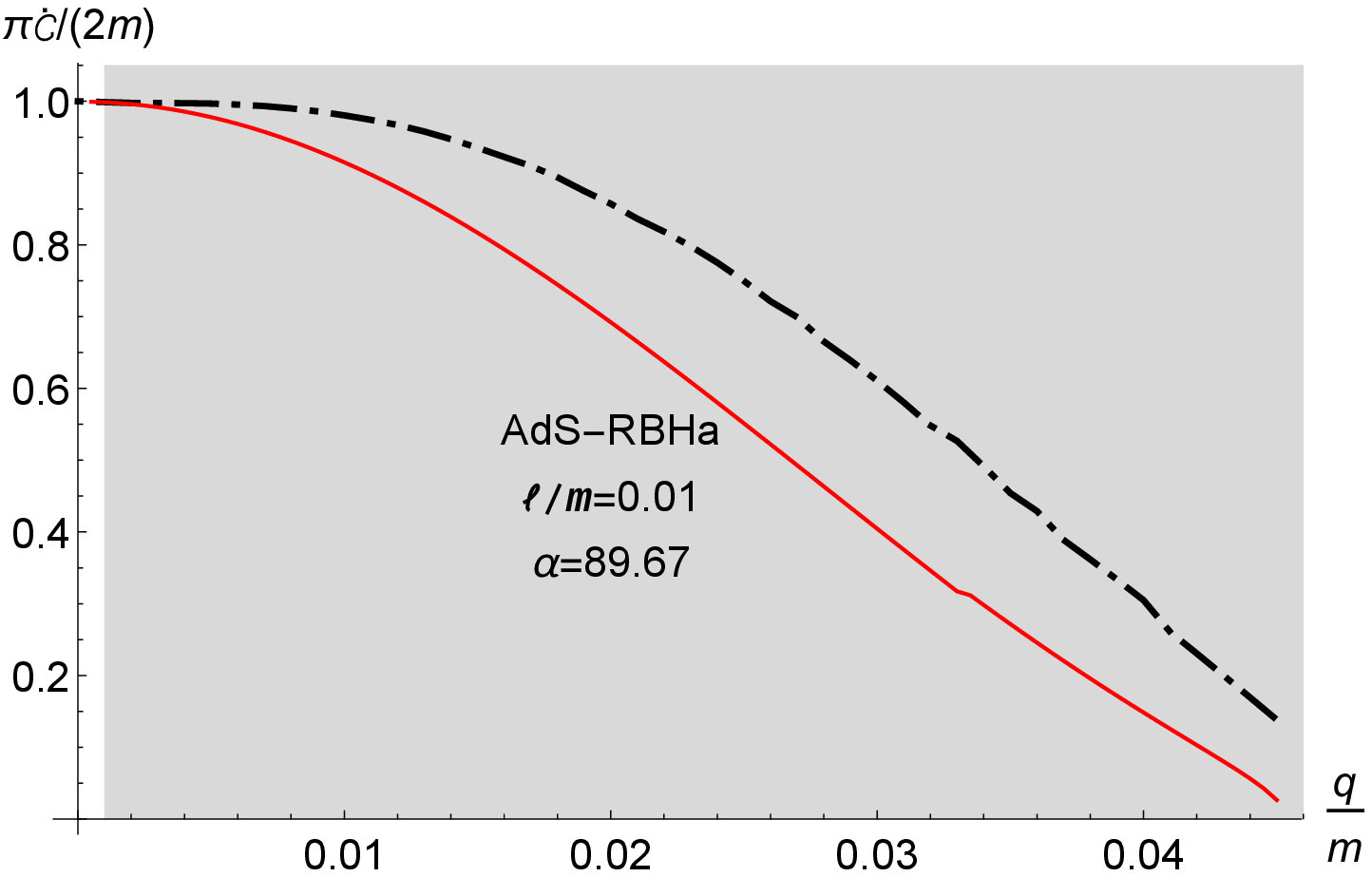}\includegraphics[width=0.3\textwidth]{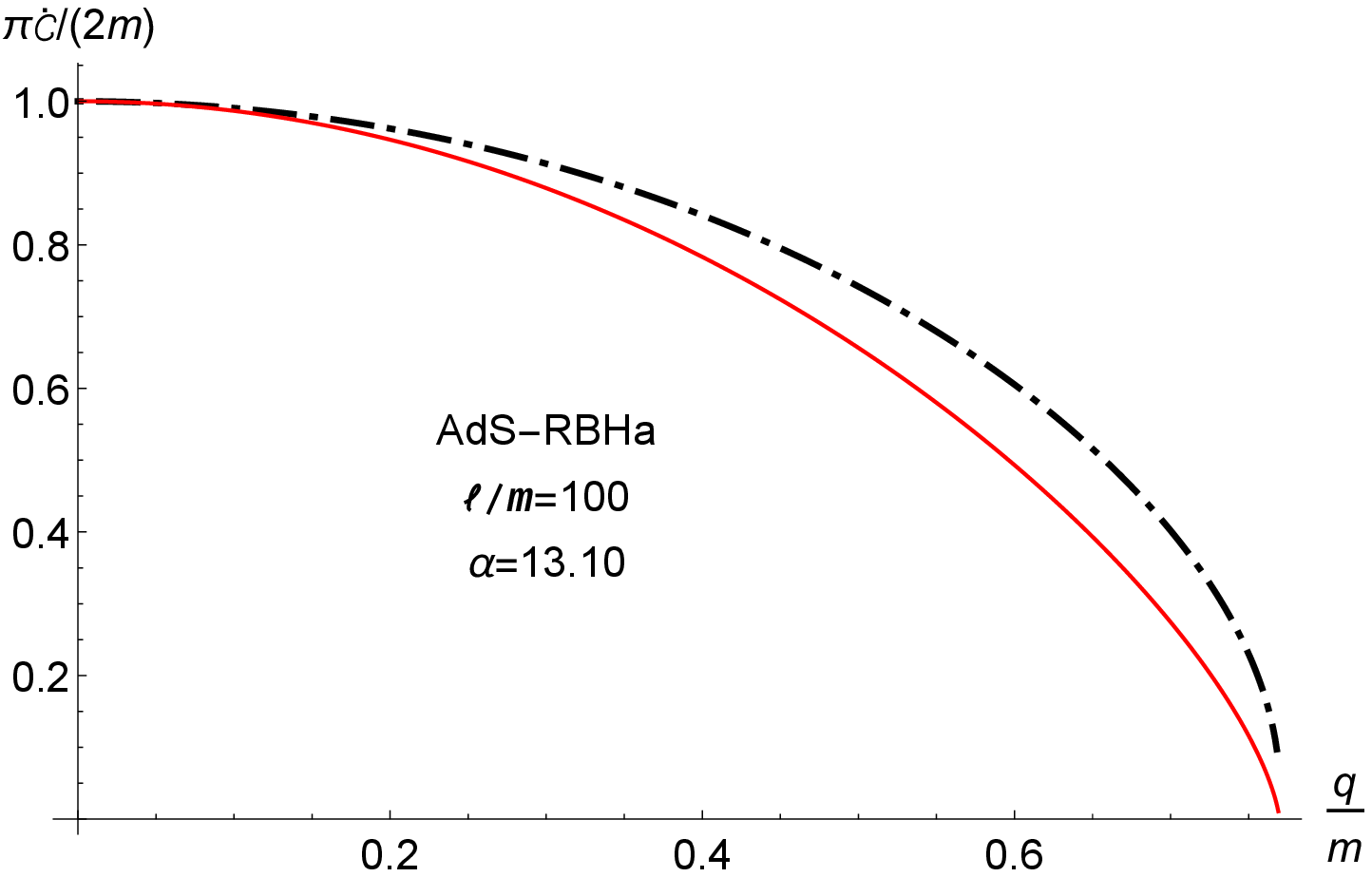}\\
\includegraphics[width=0.3\textwidth]{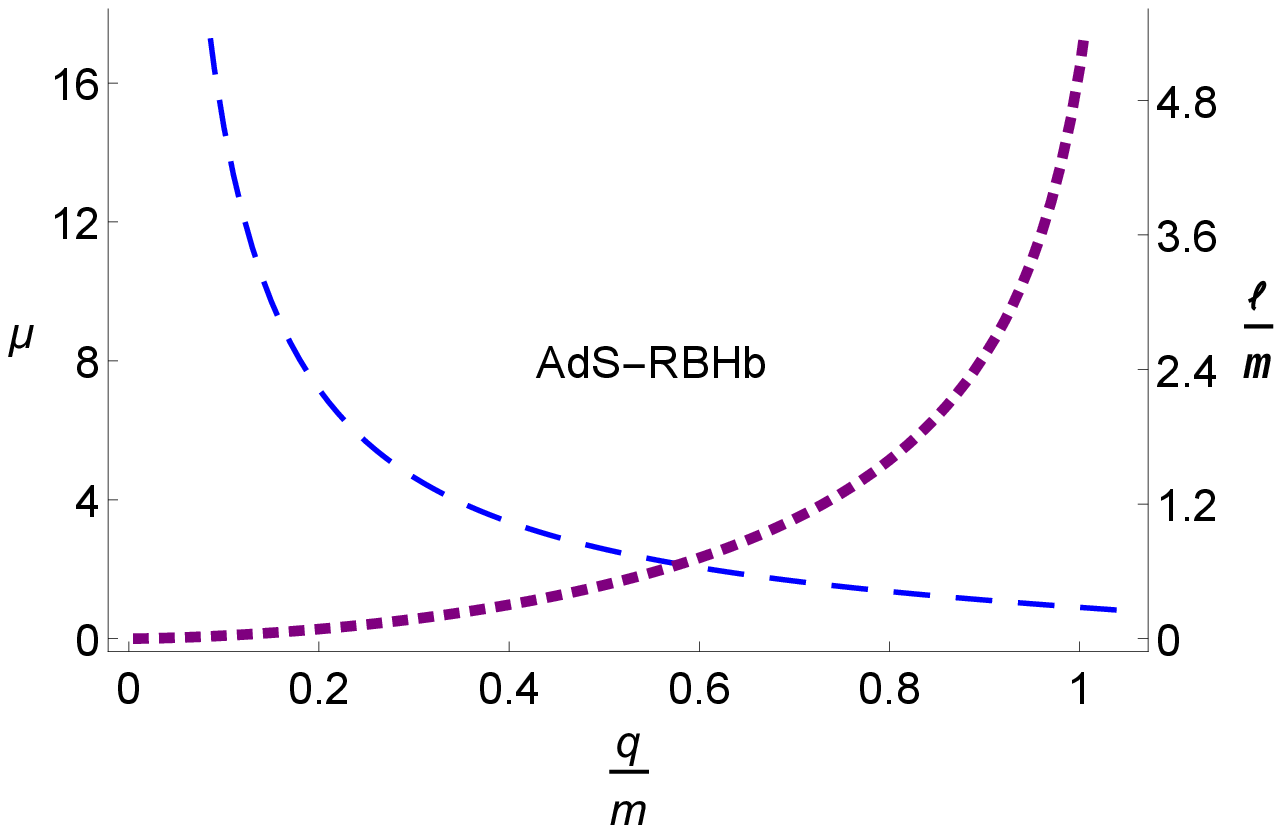}\includegraphics[width=0.3\textwidth]{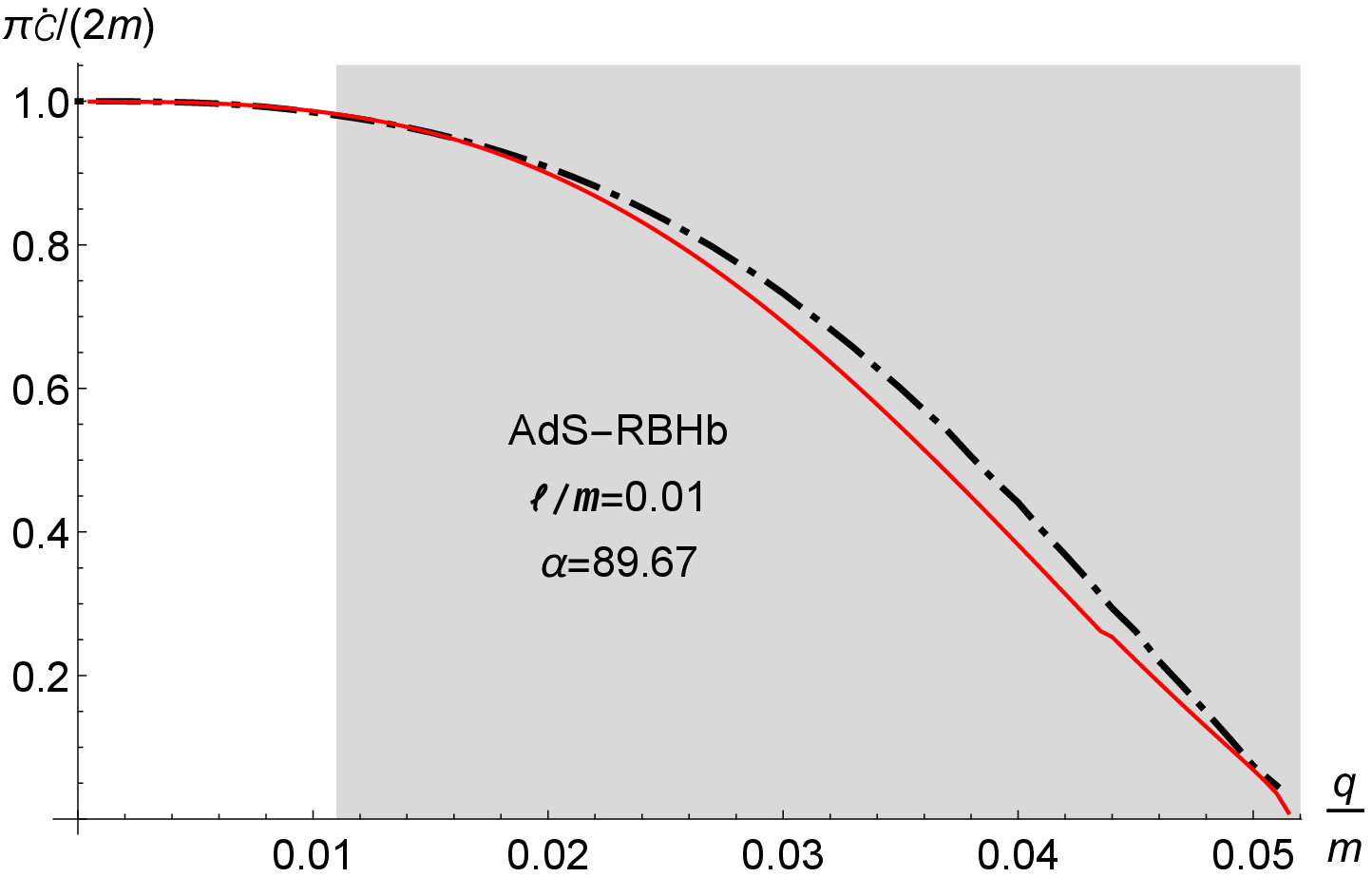}\includegraphics[width=0.3\textwidth]{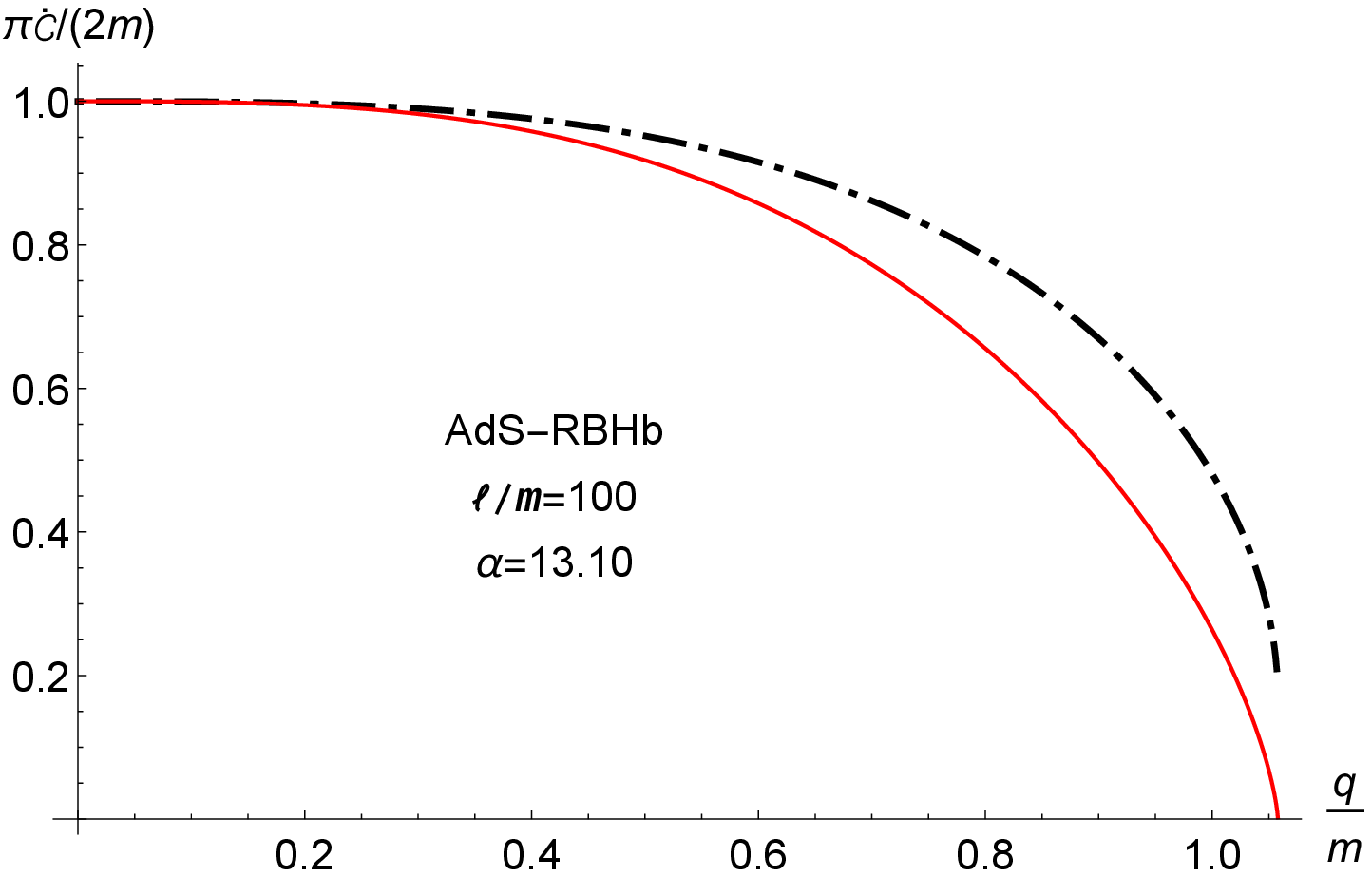}\\
\includegraphics[width=0.3\textwidth]{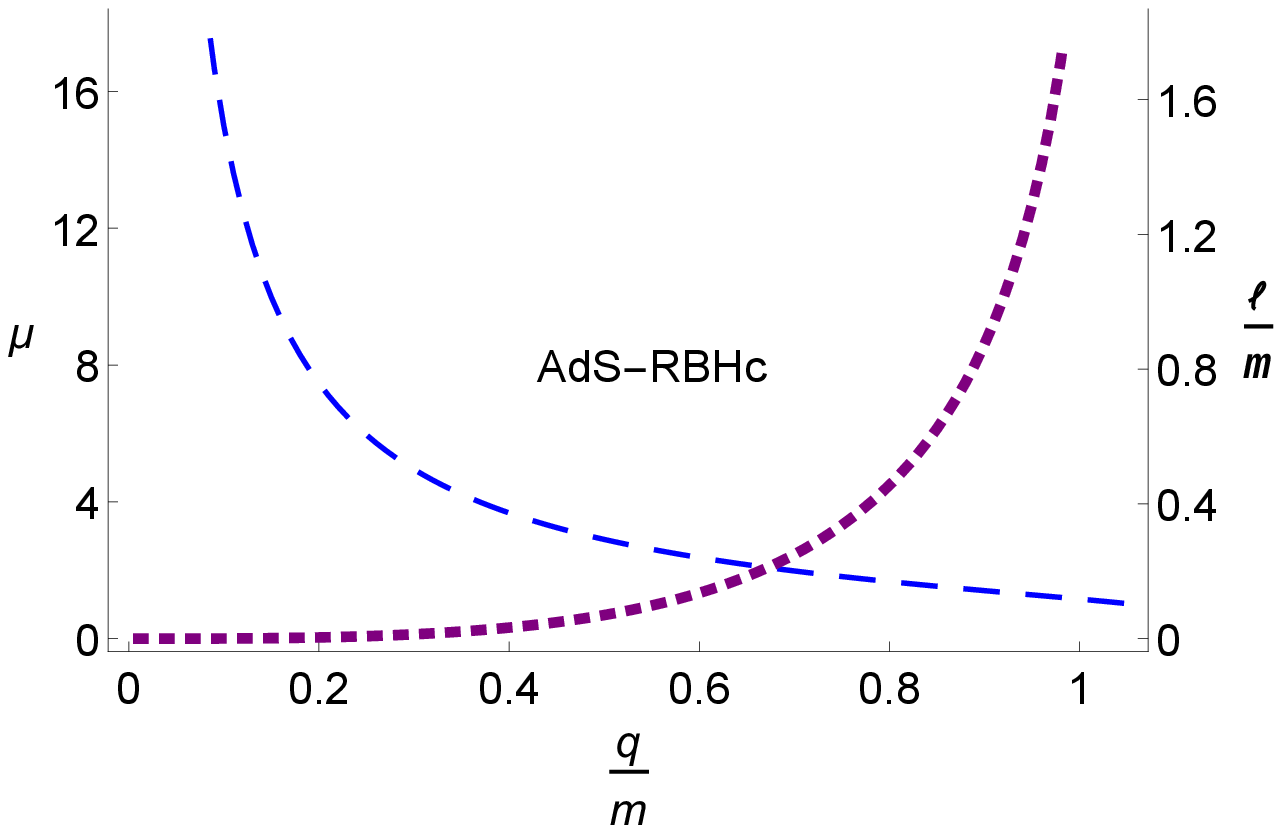}\includegraphics[width=0.3\textwidth]{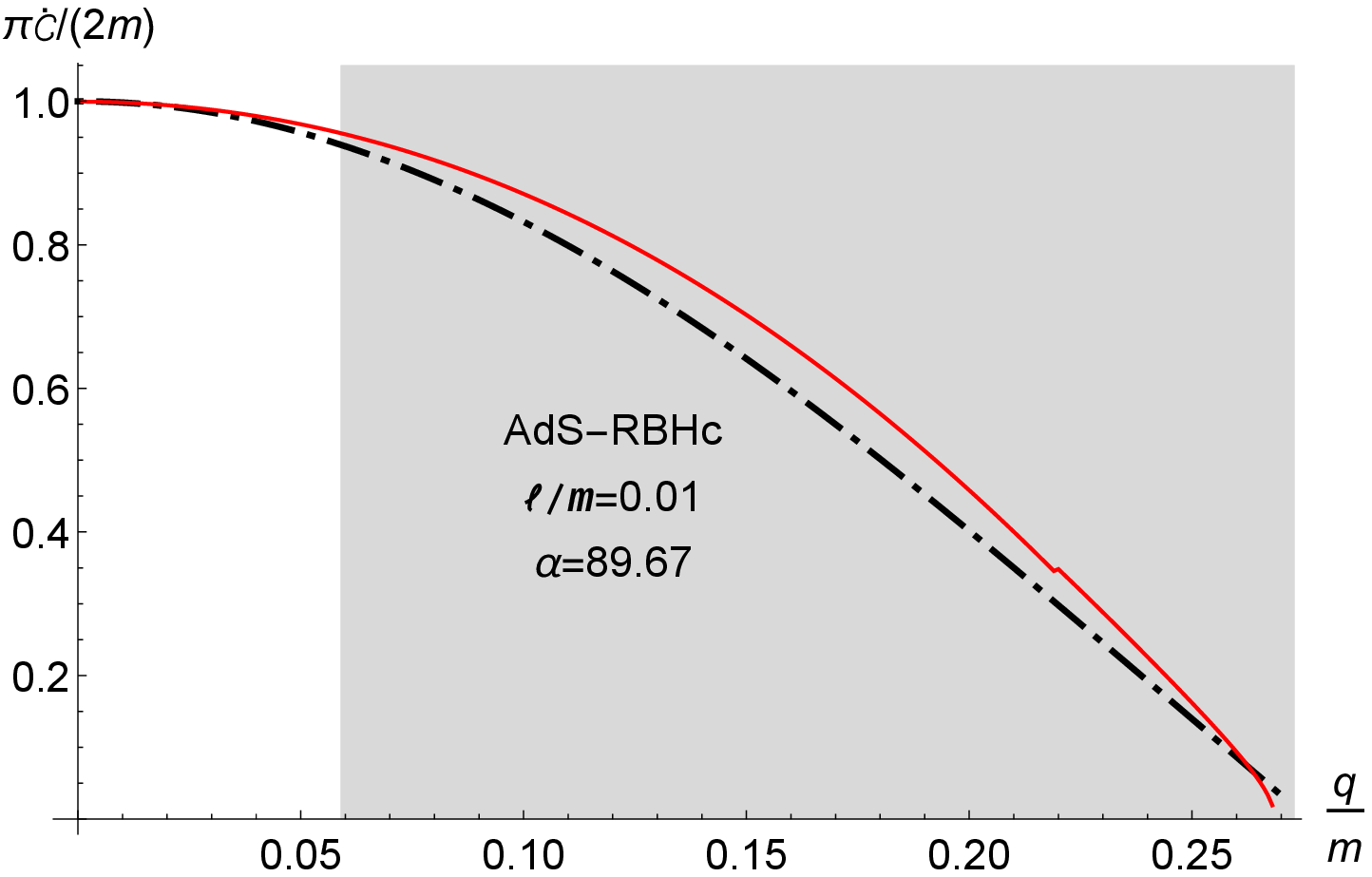}\includegraphics[width=0.3\textwidth]{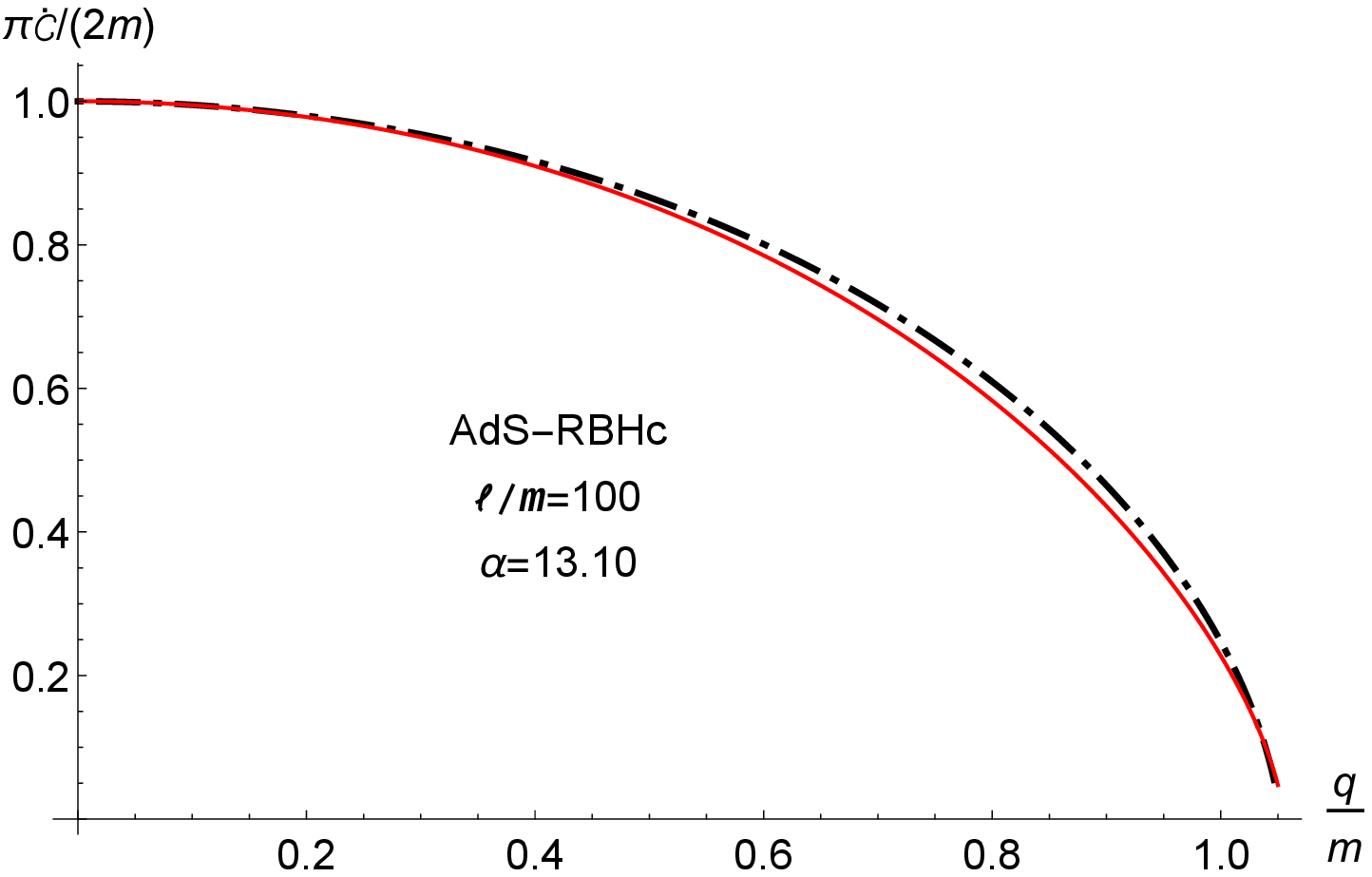}\\
\includegraphics[width=0.3\textwidth]{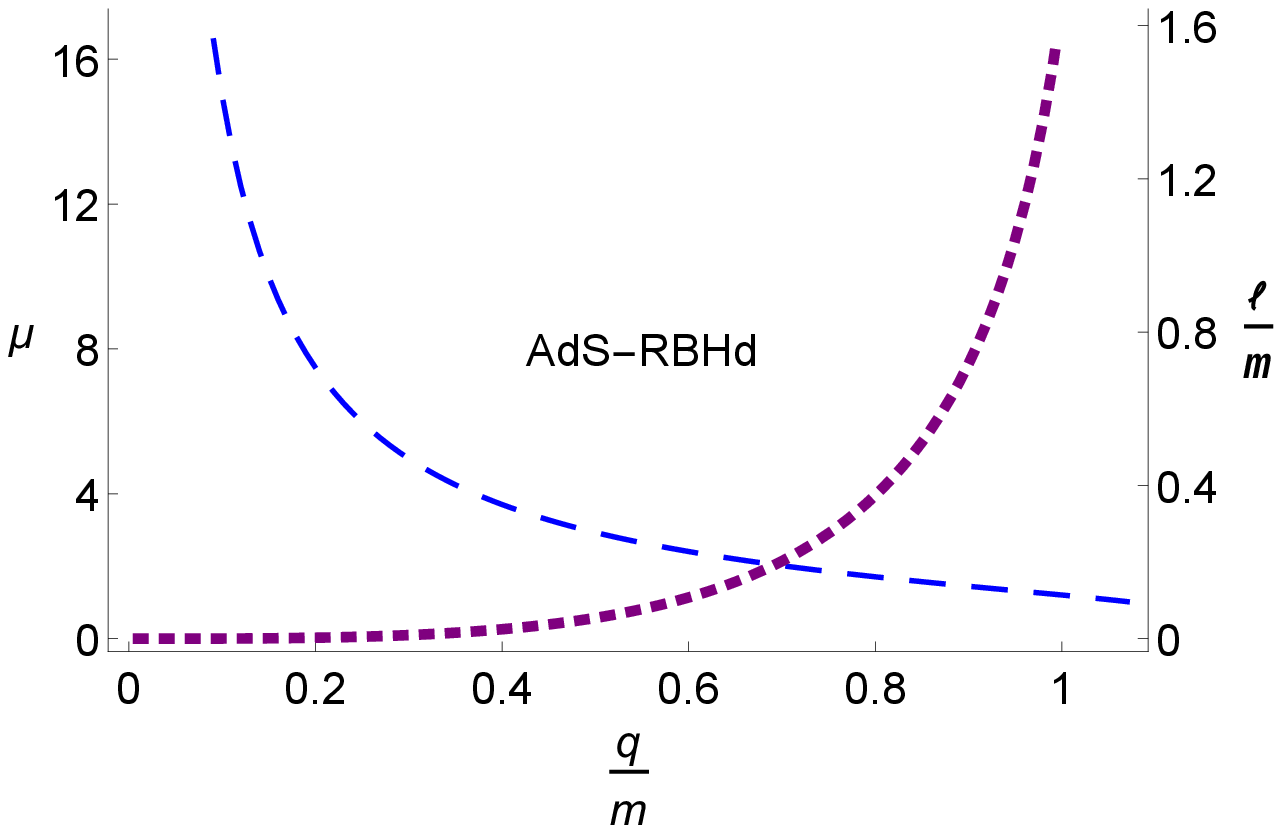}\includegraphics[width=0.3\textwidth]{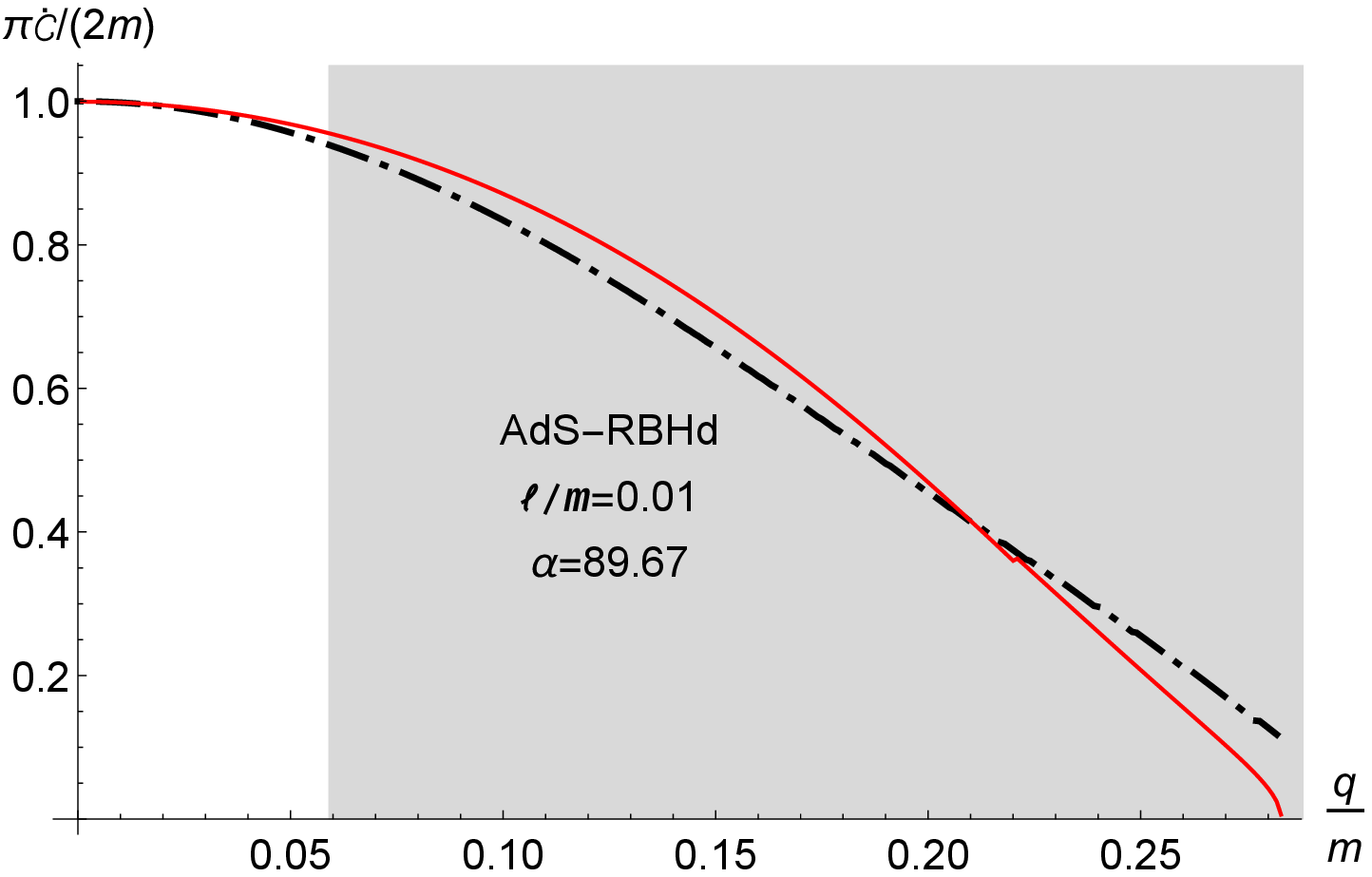}\includegraphics[width=0.3\textwidth]{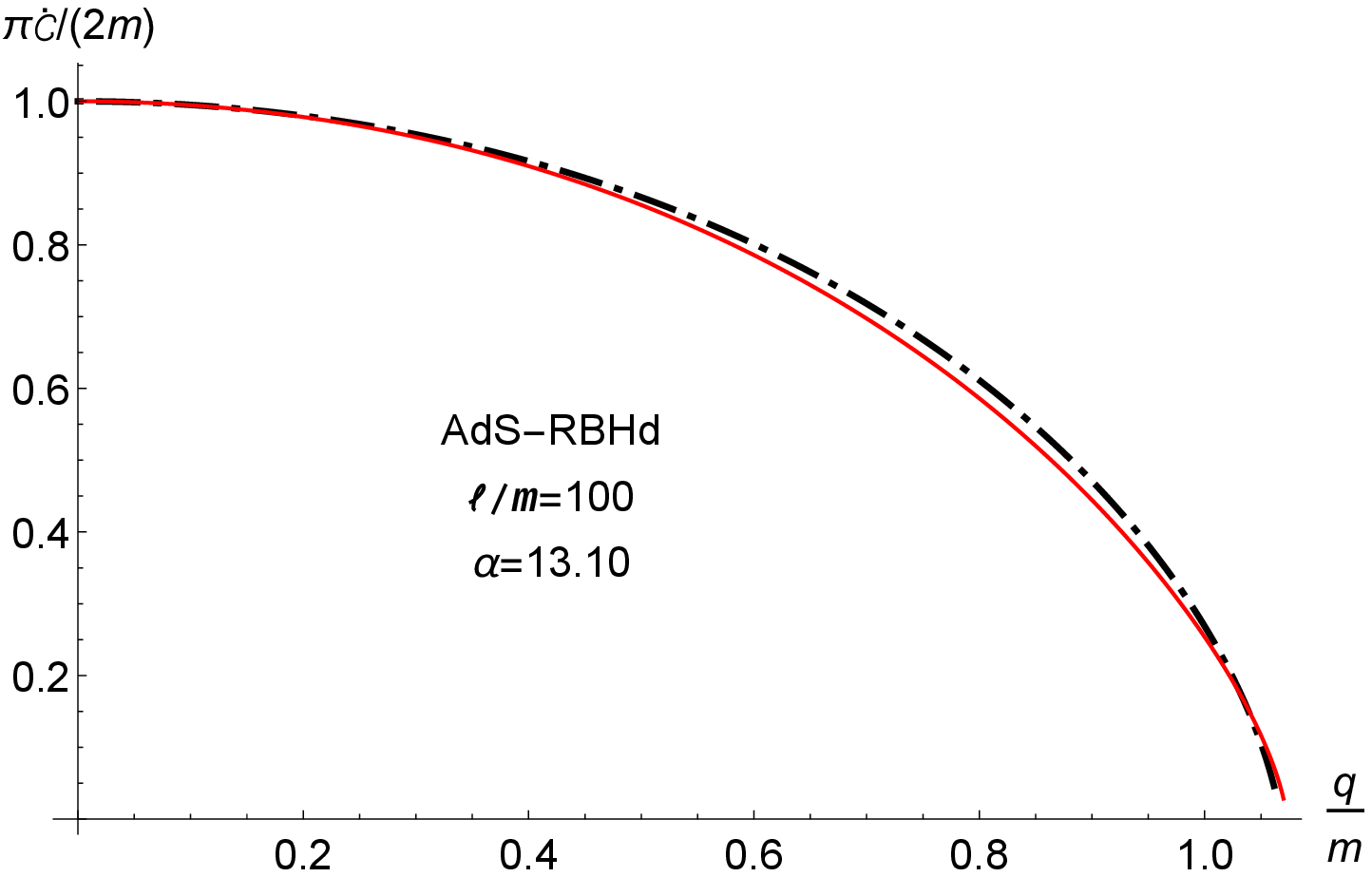}\\
\includegraphics[width=0.3\textwidth]{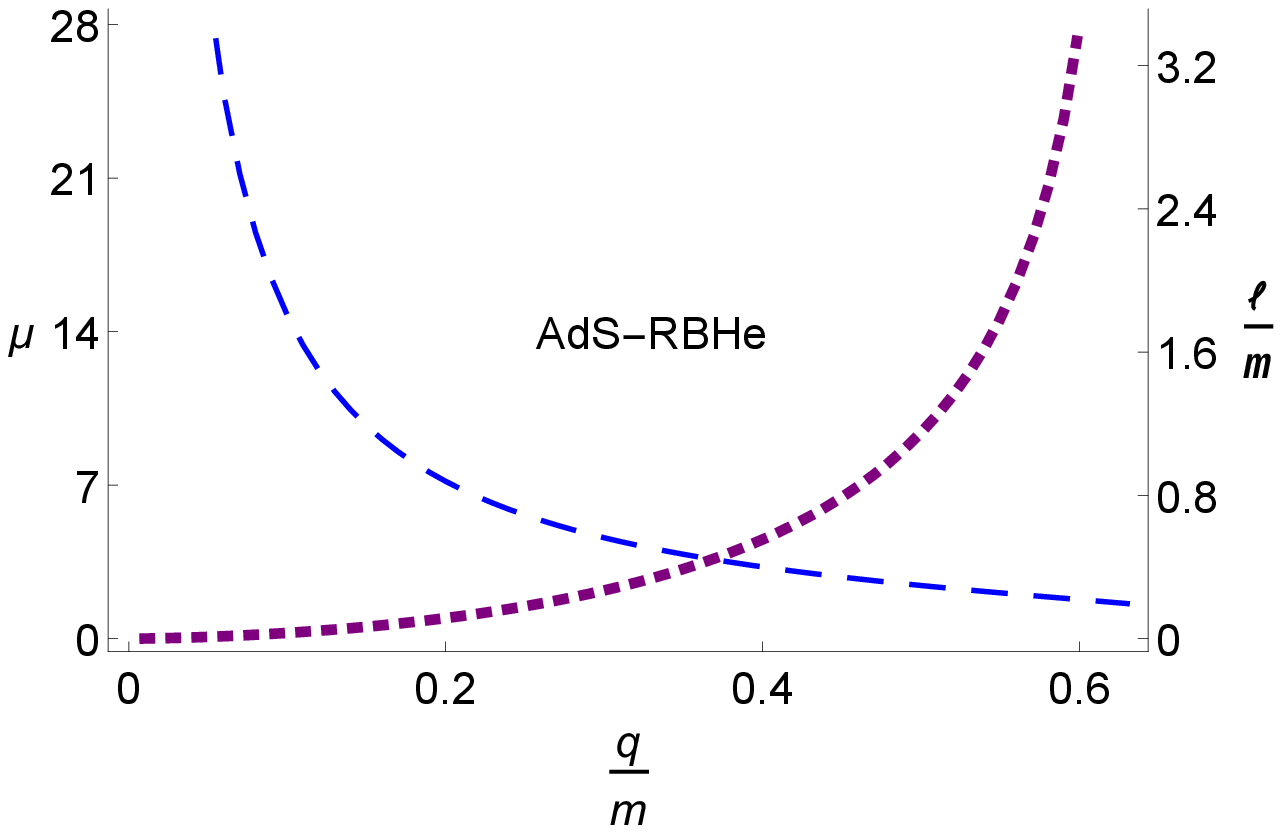}\includegraphics[width=0.3\textwidth]{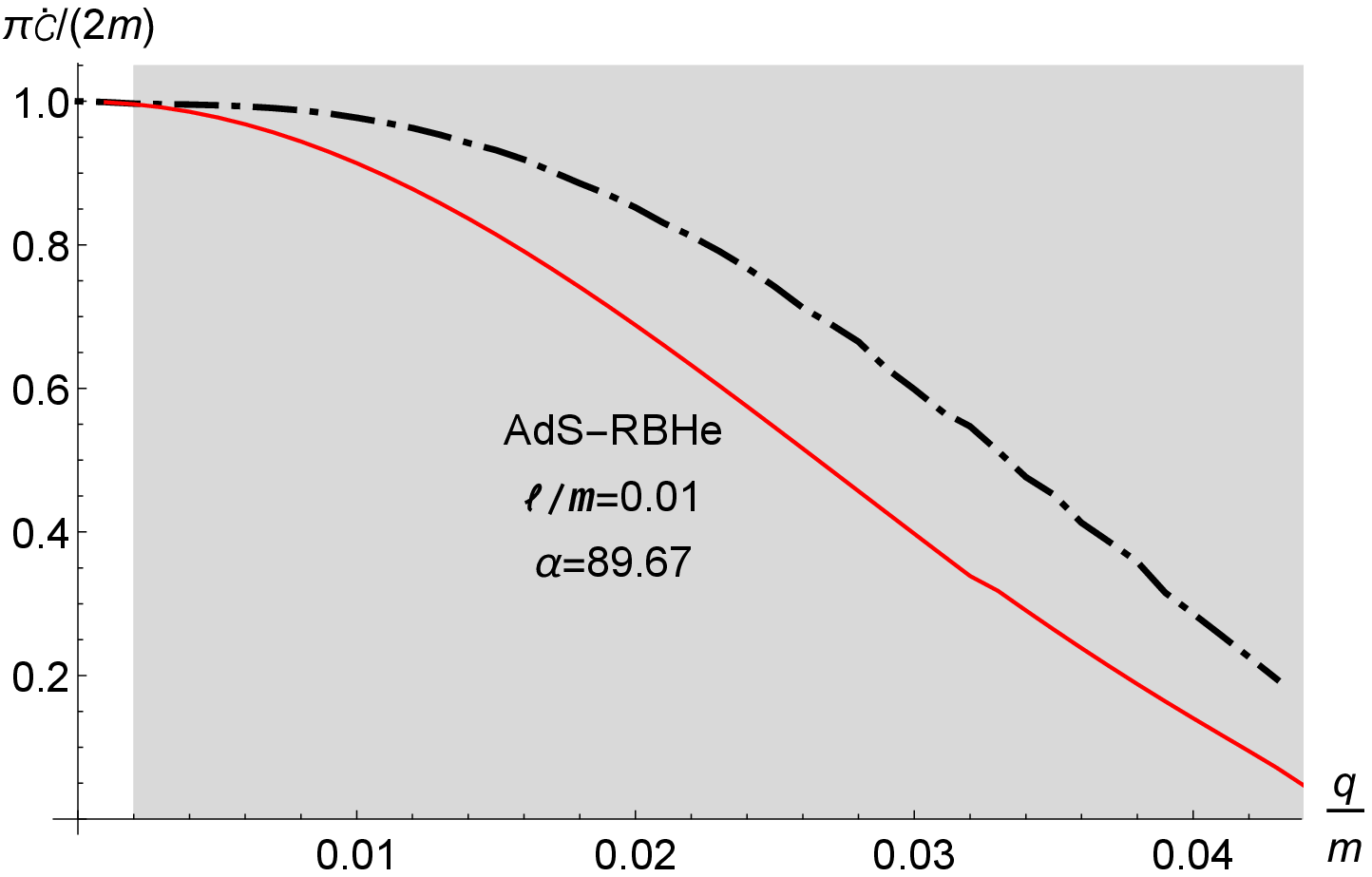}\includegraphics[width=0.3\textwidth]{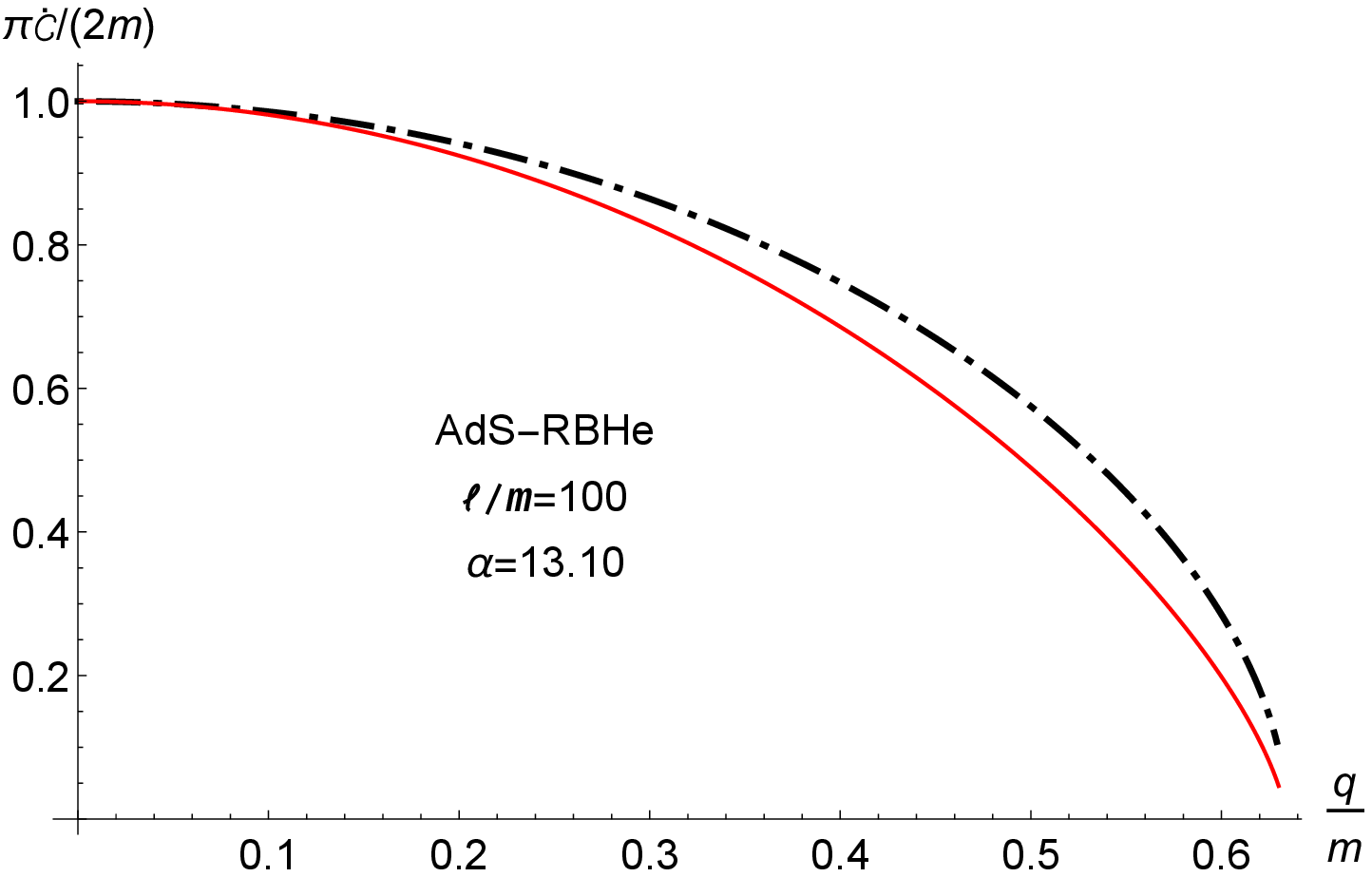}\\
\includegraphics[width=0.3\textwidth]{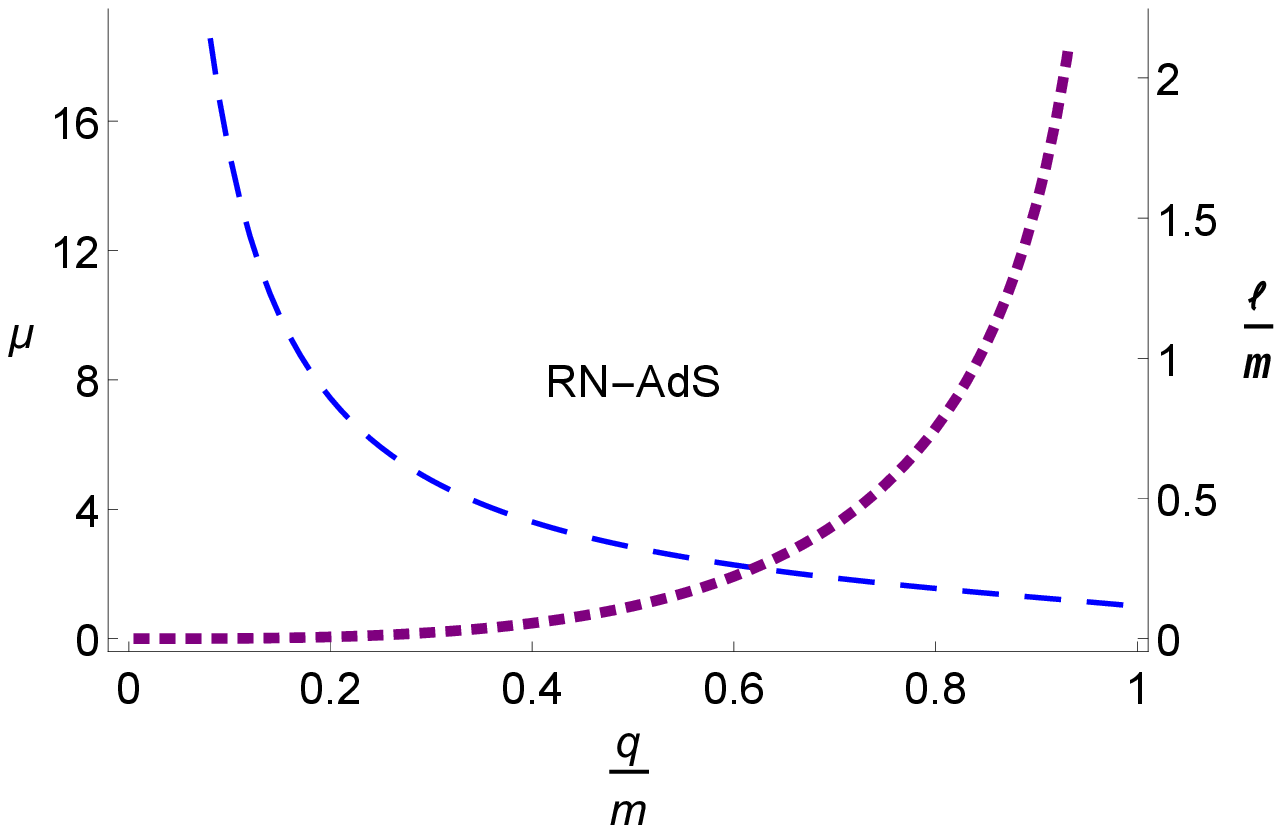}\includegraphics[width=0.3\textwidth]{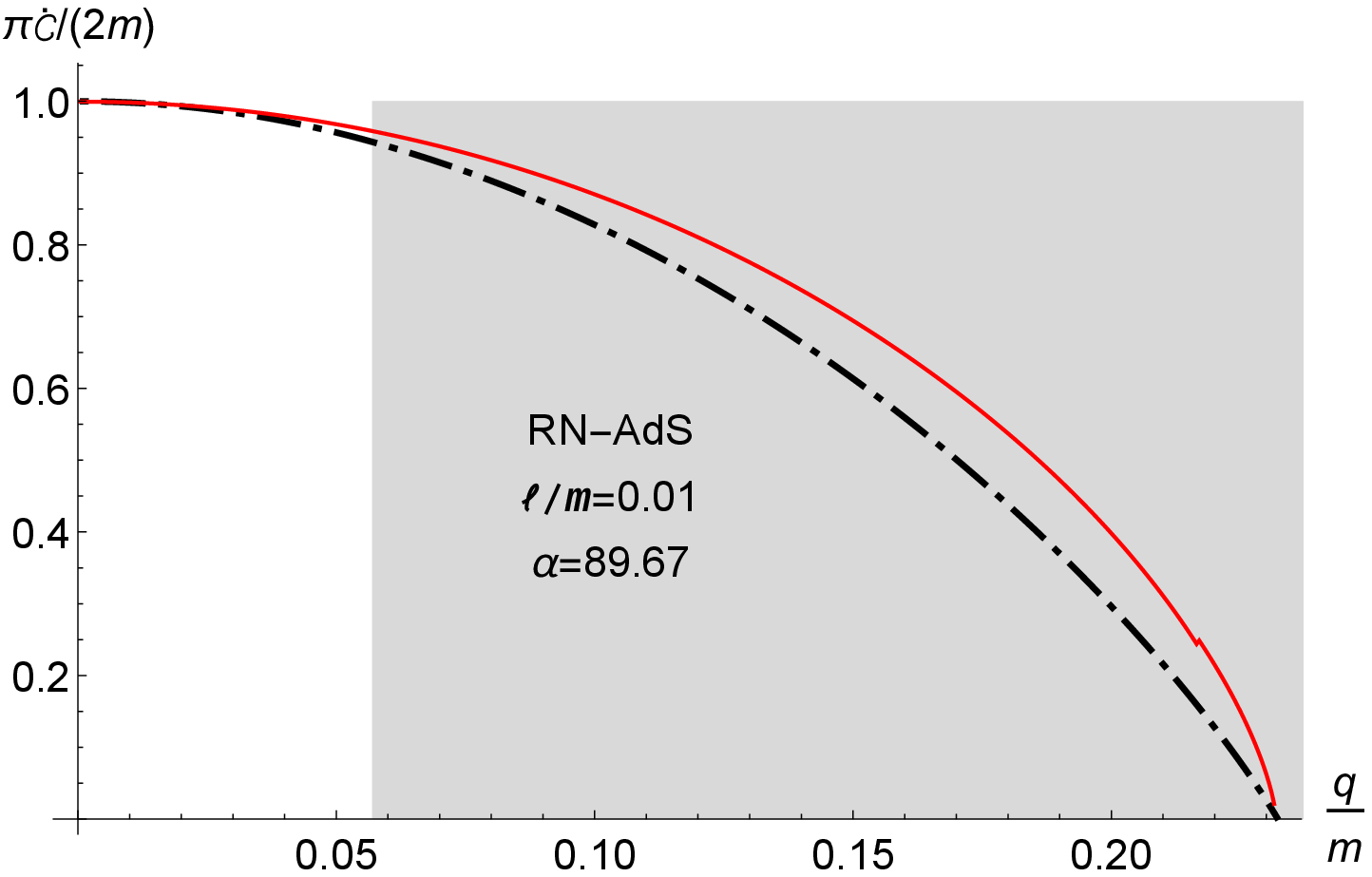}\includegraphics[width=0.3\textwidth]{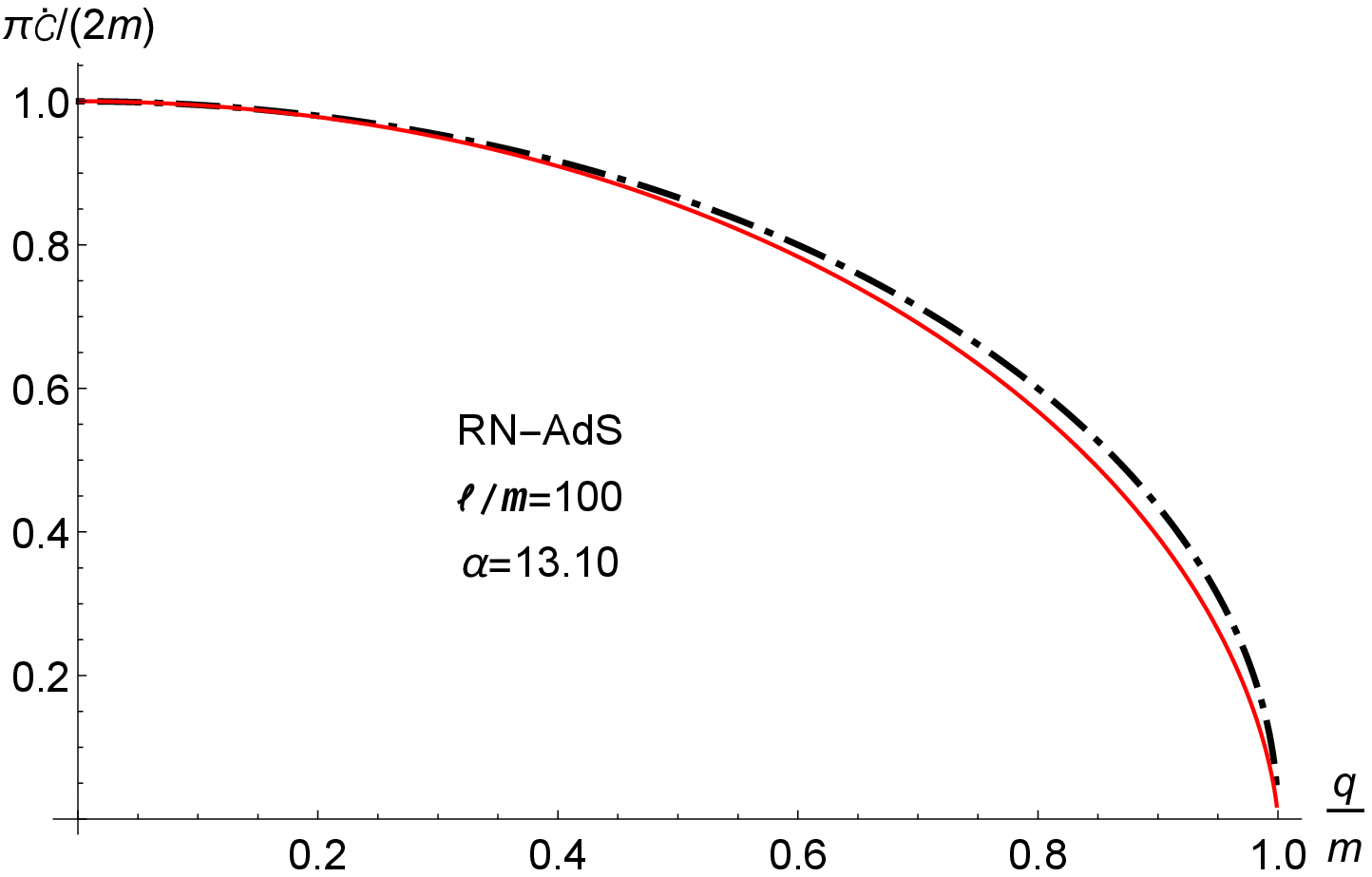}\\
\caption{(color online). In left panels, for extremal black holes, the chemical potential $\mu$ is illustrated by blue dashed curves, and $\ell/m$ by purple dotted curves as functions of $q/m$. Middle and right panels depict $\pi\dot{\mathcal{C}}/(2m)$ and its upper bound, with $\dot{\mathcal{C}}=d\mathcal{C}/dv$. Red solid lines are drawn with the complexity given by equation \eqref{CV1.1}, in which $\alpha$ is fitted by the condition $\CV=\CB$ in the Schwarzschild-AdS limit. Black dash-dotted lines display the right hand side of inequality \eqref{bd}, the normalized bound on the complexity growth rate. From top to bottom, tagged as AdS-RBHa, AdS-RBHb, AdS-RBHc, AdS-RBHd, AdS-RBHe, RN-AdS accordingly, corresponds to the black hole solutions in table \ref{tab-AdS} and solution \eqref{RN-AdS} row by row. The AdS radius $\ell$ is set to $m/100$ in all middle panels, and $100m$ in all of the right panels. In shaded parts, the ground states are extremal black holes instead of the empty space.}\label{fig-AdS}
\end{figure}

First, in the literature, most solutions of regular black holes asymptotic to a flat spacetime at $r\rightarrow\infty$ like the examples in section \ref{sect-rbh}. This obstacle is overcome in appendix \ref{app-cc}, where we demonstrate that in the Einstein gravity, from an asymptotically flat solution of the spherical static form \eqref{let}, one can get an asymptotically AdS solution by replacing the lapse function $f(r)$ with $f(r)+r^2/\ell^2$, leaving intact the electromagnetic field. The AdS radius $\ell$ is related to the cosmological constant via $\Lambda=-3/\ell^2$. Applying this prescription to solutions in table \ref{tab-rbh}, we wrote down their AdS counterparts in table \ref{tab-AdS}, tagged as AdS-RBHa, AdS-RBHb, etc. Note that AdS-RBHa and AdS-RBHb were firstly constructed in references \cite{Fan:2016rih,Fan:2016hvf}. In the neutral limit $q\rightarrow0$, all of these solutions reduce to the Schwarzschild-AdS black hole whose lapse function is
\begin{equation}\label{S-AdS}
f(r)=\frac{r^2}{\ell^2}+1-\frac{2m}{r}.
\end{equation}
Later in figure \ref{fig-AdS}, we will take this common limit to normalize the holographic complexity for asymptotically AdS regular black holes and the RN-AdS black hole.

\begin{table}[ht]
\begin{tabular}{l|c}
Lapse function & Tag \\ \hline
$f(r)=\frac{r^2}{\ell^2}+1-\frac{2mr^2}{(r^2+q^2)^{3/2}}$ & AdS-RBHa\\
$f(r)=\frac{r^2}{\ell^2}+1-\frac{2mr^2}{r^3+q^3}$ & AdS-RBHb\\
$f(r)=\frac{r^2}{\ell^2}+1-\frac{2m}{r}\left(1-\tanh\frac{q^2}{2mr}\right)$ & AdS-RBHc\\
$f(r)=\frac{r^2}{\ell^2}+1-\frac{4m}{\pi r}\left(\arctan\frac{8mr}{\pi q^2}-\frac{8m\pi q^2r}{64m^2r^2+\pi^2q^4}\right)$ & AdS-RBHd\\
$f(r)=\frac{r^2}{\ell^2}+1-\frac{2mr^2}{(r^2+q^2)^{3/2}}+\frac{q^2r^2}{(r^2+q^2)^2}$ & AdS-RBHe\\
\end{tabular}
\caption{The lapse function of some regular black holes asymptotic to the AdS spacetime at $r\rightarrow\infty$. They correspond line by line to the asymptotically flat solutions in table \ref{tab-rbh}.}\label{tab-AdS}
\end{table}

Second, in the complexity/volume duality, there is a length scale undetermined. As mentioned in reference \cite{Brown:2015lvg}, the original complexity/volume duality \cite{Stanford:2014jda} stated that the complexity of the boundary state is proportional to the spatial volume $V_{\max}$ of a maximal slice behind the horizon,
\begin{equation}\label{CV}
\mathcal{C}\sim\frac{V_{\max}}{\ell_{c}},
\end{equation}
where $\ell_{c}$ is a length parameter that has to be chosen appropriately for the configuration. For Schwarzschild-AdS black holes, $\ell_{c}$ is typically chosen as the AdS radius for large black holes and the Schwarzschild radius for small black holes \cite{Brown:2015lvg}. However, there is no universal form of $\ell_{c}$, especially for intermediate-sized black holes and charged black holes that we want to study. In fact, that was why Brown et al. turned to the complexity/action duality in references \cite{Brown:2015bva,Brown:2015lvg}.

In this paper, we will tentatively circumvent the second obstacle by exploring the possibility that
\begin{equation}\label{lc}
\ell_{c}=\sqrt{\frac{\alpha r_{+}}{T}},
\end{equation}
where $\alpha$ is a free parameter to be determined soon, $r_{+}$ is the radius of outer horizon, and the Hawking temperature $T=f'(r_{+})/(4\pi)$. This form of $\ell_{c}$ is motivated from the temperature of Schwarzschild-AdS black holes,
\begin{equation}\label{T}
T=\frac{1}{4\pi}\left(\frac{2r_{+}}{\ell^2}+\frac{2m}{r_{+}^2}\right)\sim\left\{
                                                                           \begin{array}{ll}
                                                                           \frac{3r_{+}}{4\pi\ell^2}, & \hbox{for}~r_{+}\gg\ell; \\
                                                                           \frac{1}{4\pi r_{+}}, & \hbox{for}~r_{+}\ll\ell.
                                                                           \end{array}
                                                                         \right.
\end{equation}
When taking the limits, we have made use of $\frac{r_{+}^2}{\ell^2}+1-\frac{2m}{r_{+}}=0$. Keep in mind that the large Schwarzschild-AdS black hole has $r_{+}\gg\ell$, while the small one has $r_{+}\ll\ell$. Substituting equation \eqref{T} into equation \eqref{lc}, one may check that $\ell_{c}$ tends to $2\ell\sqrt{\alpha\pi/3}$ if $r_{+}\gg\ell$ and to $2r_{+}\sqrt{\alpha\pi}$ if $r_{+}\ll\ell$. This behavior has been expected by reference \cite{Brown:2015lvg} as mentioned above, so the choice of the form \eqref{lc} for $\ell_{c}$ gives the satisfactory behavior. Inserting the ansatz \eqref{lc} into equation \eqref{CV}, we can recast the complexity/volume duality as
\begin{equation}\label{CV1.1}
\mathcal{C}\sim V_{\max}\sqrt{\frac{f'(r_{+})}{4\alpha\pi r_{+}}}.
\end{equation}

\begin{figure}
\centering
\includegraphics[width=0.3\textwidth]{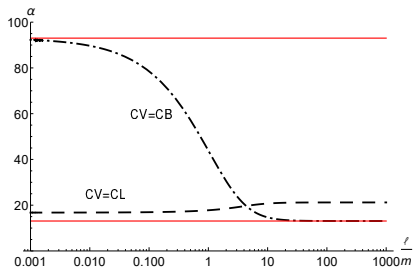}\\
\caption{(color online). The fitted value of $\alpha$ as a function of $\ell/m$ for the Schwarzschild-AdS black hole. CL, CV, CB refer to the complexity/length duality \cite{Susskind:2014rva}, the complexity/volume duality \cite{Stanford:2014jda} and the complexity growth bound conjecture \cite{Brown:2015bva,Brown:2015lvg} accordingly. The black dashed curve, $\CV=\CL$, is drawn by fitting CV to CL, i.e. $\ell_{c}\sim 4\pi A_{\min}/(TS)$. The black dash-dotted curve, $\CV=\CB$, is simulated by fitting CV to CB, i.e. $\ell_{c}\sim 2\pi^2 A_{\min}/m$. The red solid lines $\alpha=3\pi^3,(3\pi/4)^3$ are the asymptotes of $\CV=\CB$.}\label{fig-norm}
\end{figure}

The ansatz \eqref{lc} trades the length parameter $\ell_{c}$ for a dimensionless parameter $\alpha$ which is undetermined in turn. Hitherto in the literature, there are four proposals for computing the holographic complexity $\mathcal{C}$ from the bulk gravity theory: the complexity/length duality (CL) \cite{Susskind:2014rva}, the complexity/volume duality (CV) \cite{Stanford:2014jda}, the complexity/action duality (CA) \cite{Brown:2015bva,Brown:2015lvg}, and the version 2.0 of complexity/volume duality (CV2.0) \cite{Couch:2016exn}. Among them only CV has a free parameter, so we can determine $\ell_{c}$ or $\alpha$ by fitting CV to others. For example, as we have demonstrated in section \ref{sect-rbh}, CV can be made equivalent to CL if $\ell_{c}\sim 4\pi A_{\min}/(TS)$. We mark this condition as ``$\CV=\CL$'' in short. For Schwarzschild-AdS black holes, we solved this condition to get $\alpha$ as a function of $\ell/m$. The full numerical result is shown by the black dashed line in figure \ref{fig-norm}, which indicates that $\alpha$ increases monotonically as $\ell/m$ scans from zero to infinity. In the small and large $\ell/m$ limits, the analytical results are listed in the second row of table \ref{tab-norm}.

References \cite{Brown:2015bva,Brown:2015lvg} proposed that the complexity growth rate has an upper bound (CB)
\begin{equation}\label{cgb}
\frac{d\mathcal{C}}{dv}\leq\frac{2}{\pi}\left[(m-\mu q)-(m-\mu q)_{\gs}\right],
\end{equation}
where $\mu$ is the chemical potential, and $(m-\mu q)_{\gs}$ is the ground-state (smallest) value of $m-\mu q$ at fixed $\mu$. For a charged black hole with a chemical potential $\mu$, the ground state \cite{Brown:2015bva,Brown:2015lvg} is either the empty space $m=q=0$ or the extremal black hole with an equal chemical potential $\mu_{\ext}=\mu$, that is
\begin{equation}\label{gs}
(m-\mu q)_{\gs}=\min\left[0,(m-\mu q)_{\ext}\right].
\end{equation}
The subscript $\ext$ is used for quantities of extremal black holes. In references \cite{Brown:2015bva,Brown:2015lvg}, it was implicitly assumed that the ground state has the same $\ell$ as the excited states. We will take it for granted. Demanding that the Schwarzschild-AdS black holes saturate the complexity growth bound, $\CV=\CB$, one can also compute $\alpha$. The full numerical result and the partial analytical result are reported as the black dash-dotted line of figure \ref{fig-norm}, and in the third row of table \ref{tab-norm}, respectively. As $\ell/m$ varies from zero to infinity, $\alpha$ decreases monotonically.

\begin{table}[ht]
\begin{tabular}{c|l|l|l}
\multicolumn{2}{c|}{\backslashbox{Fitting condition}{Benchmark}} & $q=0,~\ell/m\rightarrow0$ & $q=0,~\ell/m\rightarrow\infty$ \\ \hline
$\CV=\CL$ & $\ell_{c}\sim 4\pi A_{\min}/(TS)$ & $\alpha=16\pi/3$ & $\alpha=27\pi/4$\\
$\CV=\CB$ & $\ell_{c}\sim 2\pi^2 A_{\min}/m$ & $\alpha=3\pi^3$ & $\alpha=(3\pi/4)^3$\\
\end{tabular}
\caption{Four asymptotic values of $\alpha$ for the Schwarzschild-AdS black hole, depending on the fitting condition and the benchmark point. CL, CV, CB refer to the complexity/length duality \cite{Susskind:2014rva}, the complexity/volume duality \cite{Stanford:2014jda} and the complexity growth bound conjecture \cite{Brown:2015bva,Brown:2015lvg} accordingly. $\CV=\CL$ means fitting CV to CL at the benchmark point, and $\CV=\CB$ indicates fitting CV to CB at the benchmark point.}\label{tab-norm}
\end{table}

In figure \ref{fig-norm}, the fitting results from $\CV=\CL$ and $\CV=\CB$ are quantitatively different, but they are still of the same order. Remarkably, although $\ell/m$ scans a large scope $(0,\infty)$, the value of $\alpha$ does not change much in the interval $(10,100)$. This suggests that, in accordance with equation \eqref{lc}, the length parameter $\ell_{c}$ scales as $\sqrt{r_{+}/T}$ up to a factor of order one. Here we have solved two fitting conditions for the Schwarzschild-AdS black holes. This is far from an exhaustive investigation, which would involve other conditions and charged black holes. We will defer it to a future work \cite{DWW}, for currently we are interested in the volume of black holes rather than the action growth rate.

The generalization of the above results to charged black holes is nontrivial, even if we restrict to the $\CV=\CB$ condition. In the general situation, $\alpha$ will be a function of both $\ell/m$ and $q/m$, and hence the black dash-dotted curve in figure \ref{fig-norm} will become a 2-dimensional surface. The ground sate of Schwarzschild-AdS black holes is always the empty space, but the ground state of charged black holes will be extremal black holes in certain parameter regions \cite{Brown:2015bva,Brown:2015lvg}. At fixed $\mu$ and $\ell$, it is tricky to work out $m_{\ext}$, $q_{\ext}$ of the extremal black hole. We explained our method in appendix \ref{app-ext} and depicted the results in figure \ref{fig-AdS} for black holes in table \ref{tab-AdS}. For comparison, in the bottom panels we presented results for the RN-AdS black hole, which is characterized by the lapse function
\begin{equation}\label{RN-AdS}
f(r)=\frac{r^2}{\ell^2}+1-\frac{2m}{r}+\frac{q^2}{r^2}.
\end{equation}

Taking the method in appendix \ref{app-ext}, we obtained the dependence of $\mu$ and $\ell/m$ on $q/m$ for extremal black holes, which are illustrated by blue dashed curves and purple dotted curves respectively in the left panels of figure \ref{fig-AdS}. Clearly, as $\mu$ increases, the extremal values of $q/m$ and $\ell/m$ decrease monotonically.

In the Schwarzschild-AdS limit \eqref{S-AdS}, the bound \eqref{cgb} is simplified to $2m/\pi$. Normalized to this limit, the bound for charged black holes takes the form \eqref{bd}, which has been plotted as the black dash-dotted lines in middle and right panels of figure \ref{fig-AdS} where $\ell/m=0.01$ and $\ell/m=100$ respectively as indicated. In these panels, the ground state is the empty space in blank parts, and becomes extremal black holes in shaded regions.

Applying the condition $\CV=\CB$ to the Schwarzschild-AdS black hole, we find $\alpha=89.67$ at $\ell/m=0.01$, and $\alpha=13.10$ at $\ell/m=100$. In principle, applied to charged black holes, the same fitting condition will produce different values of $\alpha$ for different black holes and different charges. In other words, if we fit CV to CB all the way, $\alpha$ is not simply fixed by $\ell/m$. However, to attract attention, we illustrate this point in a different way with the red solid lines in figure \ref{fig-AdS}. For all charged black holes, we fix $\alpha$ to its value in the neutral limit $q=0$, thus setting $\alpha=89.67$ for the whole red solid lines in all middle panels, and $\alpha=13.10$ for the red solid lines in right panels. Unsurprisingly, the red solid lines are mismatched with the black dash-dotted lines. But they are not mismatched too much in most cases. This reinforces that, in accordance with equation \eqref{lc}, the length parameter $\ell_{c}$ scales as $\sqrt{r_{+}/T}$ up to a factor of order one.

In the above condition $\CV=\CB$, we have expected that the complexity growth bound is always saturated by the complexity/volume duality for black hole. It is also interesting to consider the next possibility that the bound is satisfied but not always saturated by black holes, namely $\CV\leq\CB$. In reward, we could fix $\alpha$ as one universal constant for all of the examples in figure \ref{fig-AdS}. In figure \ref{fig-AdS}, CB is ``violated'' by CV when the red solid curve goes above the black dash-dotted line, slightly in some middle panels and most seriously for the RN-AdS black hole. In view of equations \eqref{CV} and \eqref{lc}, a larger $\alpha$ helps to produce a smaller $\mathcal{C}$ and cure this problem. Once we set $\alpha$ to a constant saturating DB in the most serious point, it will satisfy the bound for all. Such a universal choice of $\alpha$ is larger than $3\pi^3$, and its precise value will be reported in an extensive study \cite{DWW}.

\section{Volume outside dS horizon}\label{sect-dS}
Another distinct example of singularity-free spacetime with an event horizon is the dS spacetime, which is given by the lapse function
\begin{equation}\label{fdS}
f(r)=1-\frac{r^2}{\ell^2}.
\end{equation}
In this section, the dS radius $\ell$ is related to the cosmological constant via $\Lambda=3/\ell^2$.

As a warm-up, let us first consider the volume inside the dS horizon, which is conventionally calculated by integrate the volume measure of metric \eqref{let} at constant time $dt=0$, that is
\begin{equation}
V_{\in}=4\pi\int_0^{\ell}dr\frac{r^2}{\sqrt{f(r)}}=\pi^2\ell^3.
\end{equation}
Agreeably, the same result can be also obtained with Christodoulou and Rovelli's recipe by maximizing the integral
\begin{equation}
4\pi\int_0^{\ell}dr\frac{r^4}{\sqrt{A^2+r^4f(r)}}
\end{equation}
in the range $0<r<\ell$.

Applying Christodoulou and Rovelli's method to the outside of dS horizon, one would find a volume formally
\begin{equation}
V_{\out}=\max\left[4\pi\int_{\ell}^{\infty}dr\frac{r^4}{\sqrt{A^2+r^4f(r)}}\right].
\end{equation}
The second step is to determine $A$ by maximizing $\sqrt{-r^4f(r)}$ in the range $r>\ell$. Unfortunately, it is divergent in the limit $r\rightarrow\infty$. Even if one sets $A$ to a finite value by hand, the integrand is divergent linearly as $r\rightarrow\infty$. The estimation \eqref{Vv} is also ruined here, because deriving \eqref{Vv} relies on the condition that $r$ spans a finite region.

\begin{figure}
\centering
\includegraphics[width=0.3\textwidth]{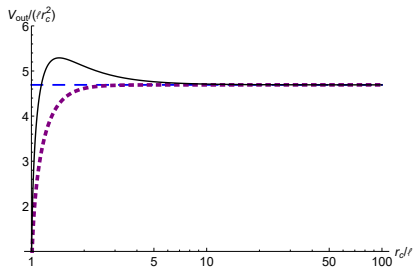}\\
\caption{(color online). The volume \eqref{VdS} outside dS horizon in terms of the cutoff radius $r_c$. The black solid line, the purple dotted line and the blue dashed line depict numerical result of the first, the third and the last lines of equation \eqref{VdS} respectively.}\label{fig-dS}
\end{figure}

To proceed, we revise this volume to a tamable form
\begin{eqnarray}
\nonumber V_{\out}&=&\lim_{r_c\rightarrow\infty}\max\left[4\pi\int_{\ell}^{r_c}dr\frac{r^4}{\sqrt{A^2+r^4f(r)}}\right]\\
&=&\lim_{r_c\rightarrow\infty}4\pi\int_{\ell}^{r_c}dr\frac{r^4}{\sqrt{r^4f(r)-r_c^4f(r_c)}}.
\end{eqnarray}
Here we have introduced a cutoff radius $r_c$ which is pushed to infinity after maximization. At large $r_c$, the volume outside dS horizon grows quadratically with cutoff radius,
\begin{eqnarray}\label{VdS}
\nonumber V_{\out}&\sim&4\pi\int_{\ell}^{r_c}dr\frac{r^4}{\sqrt{r^4f(r)-r_c^4f(r_c)}}\\
\nonumber&\sim&4\pi\int_{\ell}^{r_c}dr\frac{\ell r^4}{\sqrt{r_c^6-r^6}}\\
\nonumber&=&\frac{2\pi^{3/2}\ell r_c^2\Gamma\left(\frac{5}{6}\right)}{\Gamma\left(\frac{1}{3}\right)}-\frac{4\pi\ell^6}{5r_c^3}\,_2F_1\left(\frac{1}{2},\frac{5}{6};\frac{11}{6};\frac{\ell^6}{r_c^6}\right)\\
&\sim&\frac{2\pi^{3/2}\ell r_c^2\Gamma\left(\frac{5}{6}\right)}{\Gamma\left(\frac{1}{3}\right)}.
\end{eqnarray}
This equation involves the gamma function and the hypergeometric function. In figure \ref{fig-dS} we have numerically evaluated the first, the third and the last lines of this equation, depicted by black solid, purple dotted and blue dashed lines respectively. They coincide at large $r_c$.

In the second line of equation \eqref{VdS}, we kept only the quadratic term in the lapse function. Therefore, we expect this equation holds not only for pure dS spacetime, but also for asymptotically dS spacetime with black holes inside the dS horizon.


\section{Discussion}\label{sect-disc}
As proposed by Christodoulou and Rovelli, the volume of a black hole can be defined as the volume of maximal spacelike hypersurface bounded by the event horizon \cite{Christodoulou:2014yia}. In the present paper, we have investigated the implications of this proposal for some spacetimes with event horizons but free of singularity: asymptotically flat regular black holes, asymptotically AdS regular black holes, and the dS spacetime. The asymptotically AdS regular black holes are generated by a prescription demonstrated in appendix \ref{app-cc}, which should be useful for studying solutions to Einstein equations coupled to nonlinear electrodynamics.

Restricted to regular black holes asymptotic to the AdS spacetime at spatial infinity, in section \ref{sect-AdS} we explored the conjectures of complexity/volume duality \cite{Stanford:2014jda} and complexity growth bound \cite{Brown:2015bva,Brown:2015lvg}, assuming a length scale \eqref{lc} with a free parameter $\alpha$. Fixing the parameter $\alpha$ in the Schwarzschild-AdS limit in figure \ref{fig-AdS}, we found the bound is nearly saturated in all examples, and $\sqrt{\alpha}$ is of order one. These behaviors are encouraging. In spite of the good behaviors, we should warn the readers that the length scale \eqref{lc} is ad hoc, and our primitive exploration in section \ref{sect-AdS} relies on the assumption that the complexity growth bound is saturated for the Schwarzschild-AdS black holes. Both of these assumptions should be refined in an exhaustive study with RN-AdS and Kerr-AdS black holes as well as other geometries, which is out of the scope this paper. We plan to return to this question in a future work.

\begin{acknowledgments}
This work is supported by the National Natural Science Foundation of China (Grant No. 91536218), and in part by the Science and Technology Commission of Shanghai Municipality (Grant No. 11DZ2260700). We are grateful to Yu-Chen Ding for sharing equation \eqref{muq}. T. W. is indebted to Shi-Ying Cai for encouragement and support.
\end{acknowledgments}

\appendix

\section{A proper gesture to switch on cosmological constant}\label{app-cc}
Many solutions of regular black hole have been constructed in the literature, especially after Ayon-Beato and Garcia proposed the nonlinear-electrodynamics interpretation in reference \cite{AyonBeato:1998ub}. Most of them are of the spherical static form \eqref{let} on which we focus in this paper. As we will demonstrate in this appendix, in the Einstein gravity, there is a simple prescription to switch on cosmological constant in this class of solutions, including both regular and singular ones.

For this class of solutions, the symmetry implies that the nonvanishing components of electromagnetic strength are $F_{tr}=-F_{rt}$ and $F_{\theta\phi}=-F_{\phi\theta}$. For clarity, let us introduce notations $E\equiv F_{tr}$ and $B\equiv F_{\theta\phi}$. Furthermore, raising the indices with metric \eqref{let}, we find $F^{tr}=-F^{rt}=-E$ and $F^{\theta\phi}=-F^{\phi\theta}=B/(r^4\sin^2\theta)$. The key point is that the lapse function $f(r)$ does not appear in the expression of $F_{\mu\nu}$ and $F^{\mu\nu}$.

In the Einstein gravity, when the matter field is given by nonlinear electrodynamics, from the full action \cite{AyonBeato:1999rg,Fan:2016hvf}
\begin{equation}\label{act}
I=\frac{1}{16\pi}\int d^4x\sqrt{-g}\left[R-\mathcal{L}(\mathcal{F})\right]
\end{equation}
in which $\mathcal{F}=F_{\mu\nu}F^{\mu\nu}$, one can write down the Einstein equations
\begin{equation}
G_{\mu}^{~\nu}=2\left(\mathcal{L}_{\mathcal{F}}F_{\mu\lambda}F^{\nu\lambda}-\frac{1}{4}\delta_{\mu}^{~\nu}\mathcal{L}\right)
\end{equation}
as well as the nonlinear electrodynamic equations
\begin{equation}
\partial_{\mu}\left(\mathcal{L}_{\mathcal{F}}F^{\mu\nu}\right)+\Gamma^{\mu}_{\mu\lambda}\mathcal{L}_{\mathcal{F}}F^{\lambda\nu}=0.
\end{equation}
Substituting line element \eqref{let} and the expression of electromagnetic strength, we can put them in the form
\begin{eqnarray}\label{Eistein}
\nonumber\frac{f'}{r}+\frac{f-1}{r^2}&=&2\left(-E^2\mathcal{L}_{\mathcal{F}}-\frac{1}{4}\mathcal{L}\right),\\
\frac{f''}{2}+\frac{f'}{r}&=&2\left(\frac{B^2}{r^4\sin^2\theta}\mathcal{L}_{\mathcal{F}}-\frac{1}{4}\mathcal{L}\right),
\end{eqnarray}
\begin{eqnarray}\label{Maxwell}
\nonumber\partial_{r}\left(E\mathcal{L}_{\mathcal{F}}\right)+\frac{2}{r}E\mathcal{L}_{\mathcal{F}}&=&0,\\
\nonumber\partial_{t}\left(E\mathcal{L}_{\mathcal{F}}\right)&=&0,\\
\nonumber\partial_{\phi}\left(B\mathcal{L}_{\mathcal{F}}\right)&=&0,\\
\partial_{\theta}\left(\frac{B}{\sin^2\theta}\mathcal{L}_{\mathcal{F}}\right)+\frac{B\cot\theta}{\sin^2\theta}\mathcal{L}_{\mathcal{F}}&=&0
\end{eqnarray}
with
\begin{eqnarray}
\nonumber\mathcal{L}_{\mathcal{F}}&=&\frac{d}{d\mathcal{F}}\mathcal{L}(\mathcal{F}),\\
\mathcal{F}&=&-2E^2+\frac{2B^2}{r^4\sin^2\theta}.
\end{eqnarray}
Note that the lapse function $f(r)$ disappears in the electromagnetic equations. We also note by passing that equations \eqref{Eistein}, \eqref{Maxwell} are unchanged if one replaces $f(r)$ with $f(r)-2M/r$ where $M$ is a constant. For regular black holes, this constant is always set to zero to avoid a curvature singularity at the origin \cite{Fan:2016rih,Fan:2016hvf}.

Now we are ready to go to spacetimes asymptotic to dS/AdS at spatial infinity $r\rightarrow\infty$. For concreteness, we will focus on the AdS case with a cosmological constant $\Lambda=-3/\ell^2$. Of course, it is easy to adjust our results to the dS case by reversing the signature of $\ell^2$.

Switching on the cosmological constant in action \eqref{act} is equivalent to replacing $\mathcal{L}(\mathcal{F})$ with $\mathcal{L}(\mathcal{F})-6\ell^{-2}$. What is more, one can check that equations \eqref{Eistein}, \eqref{Maxwell} are unchanged under the following replacement
\begin{equation}
\mathcal{L}\rightarrow\mathcal{L}-\frac{6}{\ell^2},~~~~f\rightarrow f+\frac{r^2}{\ell^2},~~~~E\rightarrow E,~~~~B\rightarrow B.
\end{equation}
That is to say, {\it in the Einstein gravity, from an asymptotically flat solution of the spherical static form \eqref{let}, one can get an asymptotically AdS solution by replacing the lapse function $f(r)$ with $f(r)+r^2/\ell^2$, leaving intact the electromagnetic field.}

\section{Extremal black holes as the ground state}\label{app-ext}
As proposed in references \cite{Brown:2015bva,Brown:2015lvg}, the ground state of a black hole is not necessarily the empty space. In some parameter regions, the ground state is an extremal black hole whose chemical potential $\mu$ and AdS radius $\ell$ are the same as the ``excited state''. In both cases, to decide the ground state, one should find out the extremal black hole with given $\mu$ and $\ell$. In this appendix, we will elaborate on how to apply this idea to solutions in table \ref{tab-AdS}.

Technically, one ought to solve the system of equations
\begin{eqnarray}
&&f(r_{+})=0,\label{f}\\
&&f'(r_{+})=0,\label{df}\\
&&\mu=\mu(\ell,m,q,r_{+})\label{mu0}
\end{eqnarray}
to get the mass $m$ and the electric charge $q$ of the extremal black hole. The chemical potential $\mu=q/r_{+}$ for the RN-AdS black hole \cite{Brown:2015bva,Brown:2015lvg}. But for regular black holes, this expression cannot be used any more. If there is no magnetic field, $B=0$, which is true in our examples, it is replaced by
\begin{equation}\label{muq}
\mu q=\frac{1}{4}\left[r^2f'(r)-2rf(r)+2r\right]\Big| ^{\infty}_{r_{+}}.
\end{equation}
The formula \eqref{mu} is derived by integrating \eqref{Eistein}, \eqref{Maxwell} and then inserting the definitions of electric potential $\mu=\int Edr$ and electric charge  $4\pi q=\int \mathcal{L}_{\mathcal{F}}\ast\mathcal{F}$. Details will be presented in reference \cite{DWW}. For all of the asymptotically AdS black holes in the main text, the lapse functions behave as
\begin{equation}
f(r)\rightarrow\frac{r^2}{\ell^2}+1-\frac{2m}{r}+\hbox{negligible terms,~~~~as }r\rightarrow\infty.
\end{equation}
Here the negligible terms refer to terms that do not contribute to equation \eqref{muq}. For these black holes, equation \eqref{muq} is simplified to
\begin{equation}\label{mu}
\mu q=-\frac{1}{4}\left[r_{+}^2f'(r_{+})+2r_{+}-6m\right].
\end{equation}
Replacing equation \eqref{mu0} with \eqref{mu}, and combining it with equations \eqref{f}, \eqref{df}, we can write down a system of working equations
\begin{eqnarray}
&&2f(r_{+})-rf'(r_{+})=0,\label{ext1}\\
&&\frac{\ell}{m}=\sqrt{\frac{\ell^2}{m^2}-f(r_{+})},\label{ext2}\\
&&\mu=\frac{1}{2q}\left(3m-r_{+}\right)\label{ext3}.
\end{eqnarray}

In principle, from equations \eqref{ext1}, \eqref{ext2}, \eqref{ext3}, one can eliminate $r_{+}$ and then solve $m$, $q$ numerically. To draw figure \ref{fig-AdS}, we take an alternative method in this paper. In order to get the dependence of $\mu$ and $\ell/m$ on $q/m$, we start with a series of given values of $q/m$, and numerically solve $r_{+}$ from equation \eqref{ext1}. Substituting the obtained values of $r_{+}$ into equations \eqref{ext2}, \eqref{ext3}, we finally get a series of values of the vector $(\mu,\ell/m_{\ext},q_{\ext}/m_{\ext})$, where the subscript $\ext$ is switched on to highlight quantities for the extremal black hole. The left panels of figure \ref{fig-AdS} are plotted from this vector directly. This vector is also utilized when drawing the black dash-dotted curves in figure \ref{fig-AdS}, for which we have put equations \eqref{cgb}, \eqref{gs} into the form
\begin{equation}\label{bd}
\frac{\pi}{2m}\frac{d\mathcal{C}}{dv}\leq\left(1-\mu\frac{q}{m}\right)-\min\left[0,\frac{\ell}{m}\frac{m_{\ext}}{\ell}\left(1-\mu\frac{q_{\ext}}{m_{\ext}}\right)\right].
\end{equation}

\end{document}